\documentclass{article}

\usepackage{rotating}
\usepackage[toc,page]{appendix}

\usepackage{etoolbox}
\usepackage{amsmath}

\usepackage{appendix}
\usepackage{epstopdf}
\usepackage{placeins}  

\usepackage[colorlinks=true, linkcolor=black, citecolor=black, urlcolor=black]{hyperref}

\usepackage{subfigure}
\usepackage{threeparttable} 

\usepackage{natbib}
\bibpunct[, ]{(}{)}{;}{a}{}{,}

\usepackage{graphicx}
\usepackage{subcaption}

\usepackage{float}

\usepackage{arxiv}

\usepackage[utf8]{inputenc} 
\usepackage[T1]{fontenc}    
\usepackage{hyperref}       
\usepackage{booktabs}       
\usepackage{amsfonts}       
\usepackage{nicefrac}       
\usepackage{microtype}      
\usepackage{lipsum}		
\usepackage{graphicx}
\usepackage{natbib}
\usepackage{doi}

\usepackage{appendix}

\renewcommand{\shorttitle}{\textit{Enhancing Supply Chain Visibility with KGs and LLMs}}

\hypersetup{
pdftitle={A template for the arxiv style},
pdfsubject={q-bio.NC, q-bio.QM},
pdfauthor={David S.~Hippocampus, Elias D.~Striatum},
pdfkeywords={First keyword, Second keyword, More},
}

\title{Enhancing Supply Chain Visibility with Knowledge Graphs and Large Language Models}

\author{ Sara AlMahri \\
	Institute for Manufacturing\\Department of Engineering\\ University of Cambridge\\ Cambridge, UK \\
	\texttt{sa2104@cam.ac.uk}
	\And
	Liming Xu\\
	Institute for Manufacturing\\Department of Engineering\\ University of Cambridge\\ Cambridge, UK \\
	\texttt{lx249@cam.ac.uk}
 \And
	Alexandra Brintrup\\
	Institute for Manufacturing\\Department of Engineering\\ University of Cambridge\\ Cambridge, UK \\
 The Alan Turing Institute \\
 London, UK\\
\texttt{ab702@cam.ac.uk} 
}
\date{} 	

\UseRawInputEncoding
\begin{document}

\maketitle

\begin{abstract}
In today's globalized economy, comprehensive supply chain visibility is crucial for effective risk management. Achieving visibility remains a significant challenge due to limited information sharing among supply chain partners. This paper presents a novel framework leveraging Knowledge Graphs (KGs) and Large Language Models (LLMs) to enhance supply chain visibility without relying on direct stakeholder information sharing. Our zero-shot, LLM-driven approach automates the extraction of supply chain information from diverse public sources and constructs KGs to capture complex interdependencies between supply chain entities. We employ zero-shot prompting for Named Entity Recognition (NER) and Relation Extraction (RE) tasks, eliminating the need for extensive domain-specific training. We validate the framework with a case study on electric vehicle supply chains, focusing on tracking critical minerals for battery manufacturing. Results show significant improvements in supply chain mapping, extending visibility beyond tier-2 suppliers. The framework reveals critical dependencies and alternative sourcing options, enhancing risk management and strategic planning. With high accuracy in NER and RE tasks, it provides an effective tool for understanding complex, multi-tiered supply networks. This research offers a scalable, flexible method for constructing domain-specific supply chain KGs, addressing longstanding challenges in visibility and paving the way for advancements in digital supply chain surveillance.
\end{abstract}


\section{Introduction}

In an era where globalized and interconnected supply chains form the backbone of industries, achieving comprehensive end-to-end supply chain visibility has become a rising concern for organizations  \citet{Kalaiarasan2022TheFramework}. Supply chain visibility is "the extent to which actors within a supply chain have access
to or share information which they consider as key or useful to their operations and which they consider will be of mutual benefit" \citet{Barratt2007AntecedentsPerspective}. Visibility is viewed as not only an operational advantage, but a critical strategic asset for  risk management \citet{Emrouznejad2023SupplyTopics},  insightful decision-making \citet{Caridi2014TheModel} and operational efficiency \citet{Brusset2016DoesAgility}.

Despite its recognized importance, achieving full supply chain visibility remains a significant challenge for many organizations \citet{Zhang2011ConceptualVisibility,Caridi2014TheModel}. Limited visibility can undermine a company’s ability to proactively respond to supply chain disruptions, such as those caused by global events or geopolitical tensions. For instance, the COVID-19 pandemic starkly exposed how disruptions in distant tiers of the supply chain could ripple through to cause substantial operational and financial impact \citet{Ivanov2020ViabilityOutbreak}. Restricted visibility hinders a company's ability to ensure customer transparency regarding ethical sourcing and sustainability practices \citet{Apeji2022PrinciplesChains}. Consumers today demand more than just quality products, they require assurance that these products are sourced and manufactured ethically and sustainably. The inability to provide this transparency can damage a company's reputation and competitive standing \citet{Mollenkopf2022SupplySignals}

These challenges stem primarily from a fundamental issue: limited information sharing among supply chain partners \citet{Kumar2012InformationOverview, Prajogo2012SupplyIntegration}. This reluctance usually arises from a lack of trust among companies, as they fear that sharing information might expose their vulnerabilities or give competitors a strategic advantage \citet{Lotfi2013InformationManagement,Prajogo2012SupplyIntegration}. 

Revisiting the definition of supply chain visibility proposed by \citet{Barratt2007AntecedentsPerspective}, we understand that visibility depends fundamentally on two aspects: information sharing and information access. While challenges in information sharing are well-documented, gaining access to information through other means can also enhance visibility. This observation raises two pivotal questions: 1) Can we access information without relying on stakeholders willingness to share it? If so, 2) How can we effectively leverage this information to create valuable knowledge, thereby enhancing supply chain visibility and supporting better decision-making?

Given the demonstrated difficulties in obtaining information directly from supply chain partners, researchers have already explored alternative avenues for information access. The internet has emerged as a rich and diverse source of valuable supply chain data \citep{Wichmann2018TowardsText}. Specifically, news articles, company websites, social media, and other public sources provide rich insights into supplier-buyer relationships, production capacities, and market trends. However, web-based information presents its own set of challenges, being siloed, fragmented, and unstructured, which impedes the observation of dependencies between such decentralized data points \citep{Wichmann2020ExtractingNetworks}.

To address these challenges, researchers have explored the potential of machine learning, particularly Natural Language Processing (NLP) and Deep Learning techniques in of extracting structured information from unstructured text. Studies such as those by \citet{Kreimeyer2017NaturalReview,Li2022NeuralReview,Shickel2018DeepAnalysis,Sheikhalishahi2019NaturalReview,Wang2018ClinicalReview} have demonstrated significant advancements in this area, especially within the healthcare sector. These studies highlight how NLP can effectively convert unstructured text into structured data, thereby enhancing the ability to extract, understand, and utilize information from unstructured clinical text. 

However, only a few researchers in the supply chain field have begun adapting these methods to extract structured information from web sources. By leveraging NLP and deep learning techniques, one can interpret unstructured text and extract critical supply chain information \citet{Wichmann2018TowardsText, Wichmann2020ExtractingNetworks, Kosasih2022TowardsNetworks}. Specifically, Named Entity Recognition (NER) and Relation Extraction (RE) methods have been employed to automatically identify key elements such as company names, products, and locations, as well as to detect supplier-buyer relationships within textual data. These techniques enable the conversion of unstructured text into structured triplets (source, relation, target), facilitating the extraction of supply chain relationships.

Despite these advancements, several limitations persist. Handling domain-specific terminology remains a significant challenge, as does the ability to extend these methods to new supply chain contexts. This is mainly due to the complexity of supply chain relationships and the nuances of specific business language, as these approaches often require extensive labeled datasets for training, specially in specialized supply chain domains \citet{Kejriwal2019Domain-SpecificConstruction, Aziz2021DataNetworks, Brockmann2022SupplyGraph}, for example, tracking ethically sourced cotton or critical mineral mining.

To address these challenges, we introduce a zero-shot, Large Language Model (LLM)-driven framework that utilizes Knowledge Graphs (KGs) to enhance supply chain visibility. KGs are structured knowledge representations that organize information into interconnected nodes and edges, representing entities and their relationships, respectively \citet{EhrlingerTowardsGraphs}. This structure is particularly effective in complex environments like supply chains, as it clearly captures complex interdependencies and maps various types of relationships between different classes of entities \citet{Peng2023KnowledgeChallenges}.

Our framework leverages the extensive pre-training of LLMs, enabling them to perform NER and RE tasks specifically tailored to supply chain contexts. This is achieved through Zero-Shot Learning, a technique where the model applies learned patterns and knowledge without needing task-specific training or examples \citet{Kojima2022LargeReasoners}.

Earlier LLM models such as BERT (Bidirectional Encoder Representations from Transformers) \citet{Devlin2019BERT:Understanding}, typically require additional domain-specific data labeling and fine-tuning to perform tasks like NER and RE. This adaptation involves creating and labeling large datasets relevant to the specific domain, followed by retraining the model to understand and process the specialized terminology and relationships within that domain.

In contrast, our framework KG-LLM driven framework leverages the generalizability and zero-shot capabilities of more advanced LLMs such as OpenAI's GPT-4 \citet{OpenAI2023GPT-4Report}. These newer LLMs are pre-trained on vast and diverse datasets, enabling them to generalize across a wide range of tasks and domains without requiring additional task-specific training. Zero-shot prompting allows these models to apply their extensive pre-trained knowledge directly to new tasks, interpreting and extracting information accurately without the need for prior exposure to domain-specific examples \citet{Kojima2022LargeReasoners}.

We further prompt the LLMs to perform entity disambiguation, ensuring the uniqueness of nodes and preventing duplication, thereby preserving the integrity and consistency of the knowledge graph. 

To demonstrate the practical application of our framework, we present a  case study on improving visibility in electric vehicle supply chains. This case study illustrates how our approach can effectively track the origins of critical minerals and raw materials used in electric vehicle battery manufacturing, providing deeper insights into complex, multi-tiered supply chains and achieving levels of visibility previously difficult to attain.

Our key contributions of this work include:
\begin{enumerate}
    \item Development of a scalable methodology for constructing domain-specific supply chain knowledge graphs from diverse, publicly available data sources.
    \item Leveraging zero-shot LLMs to extract and contextualize complex supply chain relationships, extending visibility beyond tier-1 and tier-2 suppliers.
    
    \item Demonstration of the framework's effectiveness through a case study on electric vehicle manufacturers, revealing critical dependencies and alternative sourcing options for essential materials.

\end{enumerate}

\section{Literature Review}

\subsection{Knowledge Graphs in Supply Chain Management}
Knowledge graphs have emerged as a powerful tool for representing and analyzing complex relationships in various domains, including supply chain management (SCM). A knowledge graph is a structured representation of knowledge in the form of entities and relationships, providing a flexible and scalable way to integrate diverse information sources \citet{Ji2020AApplications}. Unlike supply chain networks or production networks, knowledge graphs offer greater flexibility in representing multi-dimensional relationships and can easily incorporate new information sources.
The application of knowledge graphs in SCM has shown promising results. \citet{Kosasih2022TowardsNetworks} and \citet{Deng2023ResearchManagement} demonstrated the use of knowledge graphs for supply chain risk analysis, showcasing their ability to capture complex interdependencies. \citet{Huang2019KnowledgeAnswering} proposed a knowledge graph-based approach for supply chain partner recommendation, leveraging the rich relational information encoded in the graph structure. Furthermore, \citet{Rolf2022UsingFramework} proposed using knowledge graphs and human-centric AI for reconfigurable supply chains, highlighting their potential for enhancing supply chain adaptability.

These studies underscore the potential of knowledge graphs to address the challenges of data fragmentation and complexity in supply chains. By providing a unified representation of supply chain entities and their relationships, knowledge graphs can offer enhanced visibility and support more sophisticated analysis. However, the construction of comprehensive knowledge graphs is often time-intensive and challenging, particularly when dealing with large-scale supply chains.

\subsection{Natural Language Processing for Supply Chain Mapping}

To address the challenges of manual knowledge graph construction, researchers have explored the use of NLP techniques for automated supply chain mapping. Supply chain mapping involves creating a visual representation of the relationships and flows between different entities in a supply chain \citet{MacCarthy2022MappingHow}. Two key NLP techniques, NER and RE, have shown promise in automating this process by extracting relevant information from unstructured text sources.

NER is a subfield of NLP that focuses on identifying and classifying specific data points, known as named entities, within a text \citet{Wang2023GPT-NER:Models}. These entities can include names of people, organizations, locations, dates, and other significant terms. NER works by using algorithms trained on labeled datasets to detect and categorize these entities, transforming unstructured text into structured data \citet{Sun2018AnRecognition}.

RE, on the other hand, is an NLP technique that identifies and categorizes the connections between entities mentioned in a text \citet{Wadhwa2023RevisitingModels}. This process helps in understanding how different entities are related to each other, such as "works at", "located in" or "founded by."

BERT, introduced by \citet{Devlin2019BERT:Understanding}, revolutionized the field of NLP with its bidirectional training approach. BERT is a type of LLM designed to understand the context of words in a sentence by looking at the words that come before and after. This model is pre-trained on a large corpus of text and can be fine-tuned for specific tasks, making it highly effective when adapted to the target domain. Building on this work, \citet{Liu2019RoBERTa:Approach} proposed RoBERTa, an optimized version of BERT with improved performance across various NLP tasks.

Researchers have explored the application of different NLP methods in supply chain mapping. For example, \citet{Wichmann2018TowardsText} proposed using NER and RE methods to generate supply chain maps from online unstructured text sources. Their approach demonstrated the feasibility of automatically extracting supply chain relationships from news articles. Similarly, \citet{Yamamoto2017CompanyStructure} developed a method for extracting company relationships from web news articles to analyze industry structure. In another study, \citet{Wichmann2020ExtractingNetworks} explored using deep learning to automatically extract buyer–supplier relations from natural language text.

While these studies demonstrate the potential of machine learning in supply chain mapping, they also highlight certain limitations. These methods often require extensive pre-training and labeled datasets, which can be challenging to obtain in specialized domains like SCM. Moreover, the complexity of supply chain relationships and the nuances of business language can pose challenges for automated extraction methods.

\subsection{Large Language Models \& Knowledge Graphs}

While BERT is considered an LLM, it is primarily used for understanding and processing text, specifically tailored for tasks like NER and RE through fine-tuning \citet{Devlin2019BERT:Understanding}. The key advantage of newer LLMs lies in their generalizability: they can understand and generate human-like text, perform complex tasks across various domains, and adapt quickly to new tasks with little to no additional training \citet{Kojima2022LargeReasoners}. 
The emergence of more advanced LLMs has marked a significant advancement in NLP. LLMs, such as GPT-3 introduced by \citet{Brown2020LanguageLearners}, have demonstrated remarkable capabilities across various NLP tasks, including Zero-Shot and Few-Shot learning. Zero-shot learning refers to the ability of a model to perform a task it has never been explicitly trained on by leveraging its pre-existing knowledge and patterns learned during its extensive pre-training phase \citet{Kojima2022LargeReasoners}. This means the model can understand and respond to new types of tasks or questions without having seen any specific examples of them before.

Few-shot learning, on the other hand, allows the model to adapt to new tasks with only a few examples or instances of the task provided. In this scenario, the model uses a limited number of task-specific examples to understand the new task better, enhancing its performance without requiring a large, labeled dataset for training \citet{Brown2020LanguageLearners}.

These capabilities enable them to efficiently handle tasks in specialized domains like supply chain management. Their ability to perform zero-shot and few-shot learning means they can quickly adapt to domain-specific terminology and relationships without the need for extensive task-specific data labeling and fine-tuning

Recent research has explored the potential of LLMs in automating knowledge graph construction. 
\citet{Faria2023QuestionGPT-3} demonstrated the use of GPT-3 for zero-shot knowledge graph completion, achieving competitive performance with minimal task-specific training. \citet{Zhu2023LLMsOpportunities} showed that LLMs can effectively extract triples from unstructured text for knowledge graph construction. Furthermore, several studies highlighted the superior performance of LLMs as they can achieve near state-of-the-art performance with only Few-Shot Prompting, roughly equivalent to existing fully supervised models on both NER and RE tasks \citet{Wadhwa2023RevisitingModelsb}. Additionally, studies report LLM's superior ability in situations where the amount of training data is extremely scarce, as they can perform significantly better than supervised models \citet{Wang2023GPT-NER:Modelsb}.

While LLMs have been applied to various domains, their potential in supply chain management, particularly for enhancing visibility, remains largely unexplored. \citet{Li2023LargeOptimization} conducted one of the first studies exploring the use of LLMs for supply chain optimization, demonstrating their potential for bridging the gap between supply chain automation and human comprehension.

\subsection{Research Gaps and Opportunities}
Despite notable progress in the use of NLP and knowledge graphs in enhancing supply chain visibility, significant challenges remain:

\begin{itemize}

\item There is a noticeable lack of flexibility in integrating diverse and heterogeneous data sets, which is crucial for comprehensive supply chain visibility.

\item Earlier supply chain mapping approaches require extensive labeled data and further fine-tuning to effectively manage supply chain-specific terminology.  

\item The potential for enhancing supply chain visibility by integrating LLMs with knowledge graphs remains largely untapped and requires further exploration.
\end{itemize}

In this paper, we address these identified gaps and present our contributions through the following sections. In \textbf{Section 3: Methodology}, we discuss our proposed KG-LLM-driven framework in detail, outlining the steps involved in constructing it to enhance supply chain visibility. \textbf{Section 4: Experimental Studies} presents our evaluation method for the developed framework, validates its performance, and provides results to quantitatively assess the framework's accuracy. Additionally, we include a case study to qualitatively demonstrate the potential of this framework to enhance supply chain visibility. Finally, in \textbf{Section 5: Conclusion}, we summarize our observations, discuss the limitations of our study, and suggest areas for future work.

\section{Methodology}

In this section, we begin with an overview of the KG-LLM framework designed to improve supply chain visibility. We then delve into the detailed stages of the framework, covering dataset collection and pre-processing, as well as the construction of the knowledge graph. This includes three key tasks: NER, RE, and entity disambiguation.

\begin{figure} [htbp]
    \centering
    \includegraphics[width=1\linewidth]{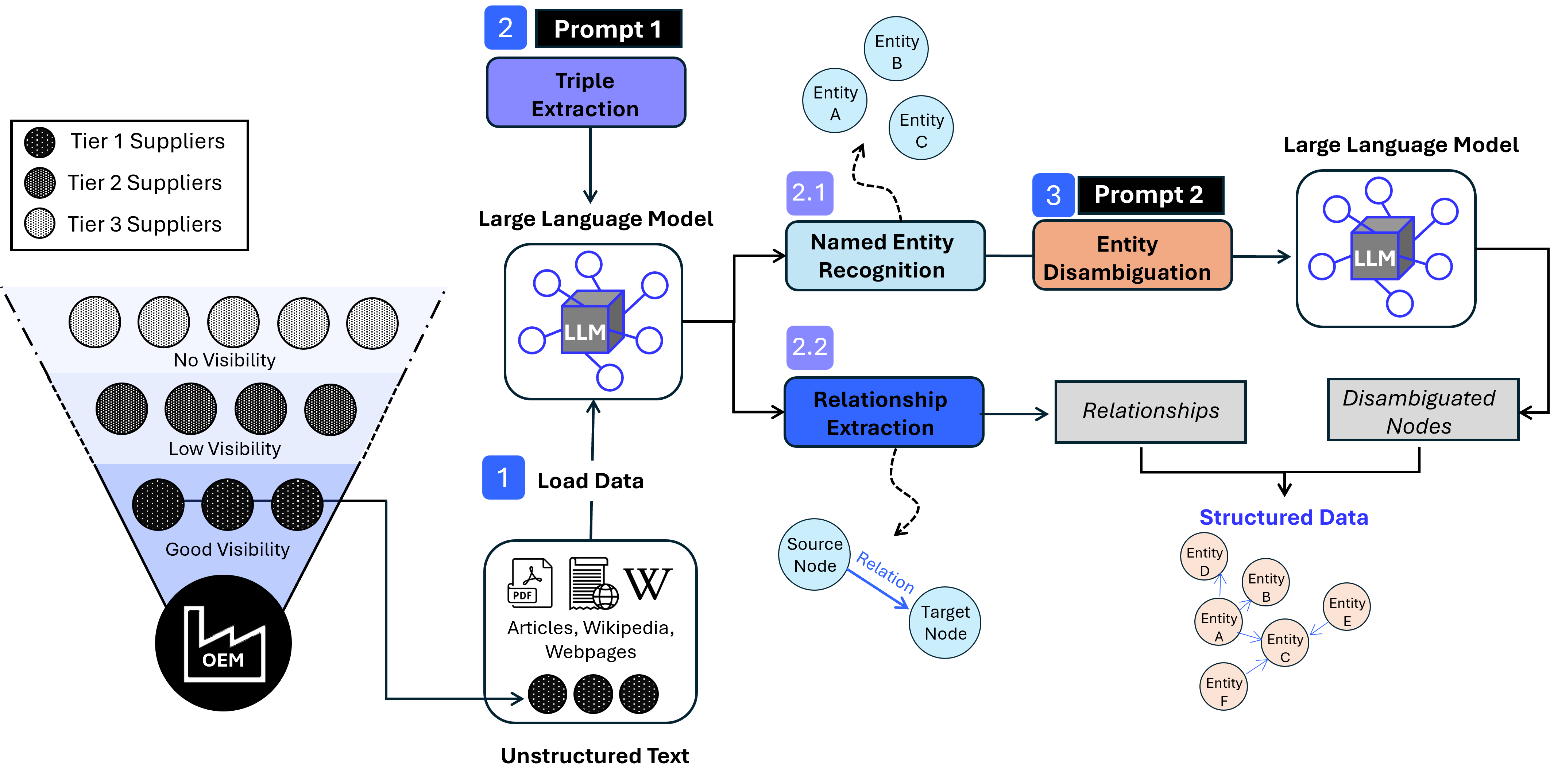}
    \caption{Diagram of the Knowledge Graph-Large Language Model Framework Developed for Enhancing Supply Chain Visibility}
    \label{framework_overview}
\end{figure}

\subsection{LLM-driven Knowledge Graph Framework for Enhanced Supply Chain Visibility}

As shown in Figure \ref{framework_overview}, we introduce a KG-LLM driven framework. Our approach addresses the critical challenge of extracting, structuring, and centralizing the vast amount of supply chain information dispersed across the web.
While this information is crucial for understanding extended supplier networks, its fragmented state severely limits its utility. 
To overcome these limitations, we develop a framework that combines the structural advantages of knowledge graphs with the advanced NLP capabilities of LLMs. We employ a zero-shot learning approach, leveraging the contextual understanding and language processing capabilities of state-of-the-art GPT-4 LLM by OpenAI \citet{OpenAI2023GPT-4Report}. This integration allows for the automated construction of comprehensive, interconnected representations of supply chain ecosystems. As shown in Figure \ref{framework_overview}, our framework begins with systematically identifying and aggregating data from various unstructured sources such as online web pages, articles, and Wikipedia entries, forming the foundation for extracting the extended supply network. We then develop two main prompts for different stages in the framework: the first facilitates the extraction of entities and relationships, essential for NER and RE, while the second ensures precise distinction of entities during the entity disambiguation stage. Using the LLM, we extract significant data points like company names, locations, and product details in the NER process. The RE process determines the relationships between identified entities, articulating interactions and dependencies critical for constructing the links between nodes in the knowledge graph. Finally, entity disambiguation resolves ambiguities among entities to ensure each node in the knowledge graph is unique, maintaining the graph's integrity and consistency. In the following sections, we discuss each step in more details.

\subsection{Data Collection \& Pre-Processing}
The initial stage involves systematically identifying and aggregating data from various unstructured web sources. It is crucial to emphasize that the selection of data sources is critical to the integrity and credibility of the extracted insights. We note that when implementing this framework, the sources should be carefully evaluated for relevance, accuracy, and reliability to ensure the quality of the resulting knowledge graph.
It is also important to note that this framework provides static information and knowledge. While exploring temporal knowledge graphs is a promising avenue for future research, it is beyond the scope of this study. Our current focus is on establishing a baseline framework for integrating LLMs and knowledge graphs to enhance supply chain visibility.

\begin{figure}[hbt]
    \centering
    \includegraphics[width=0.9\linewidth]{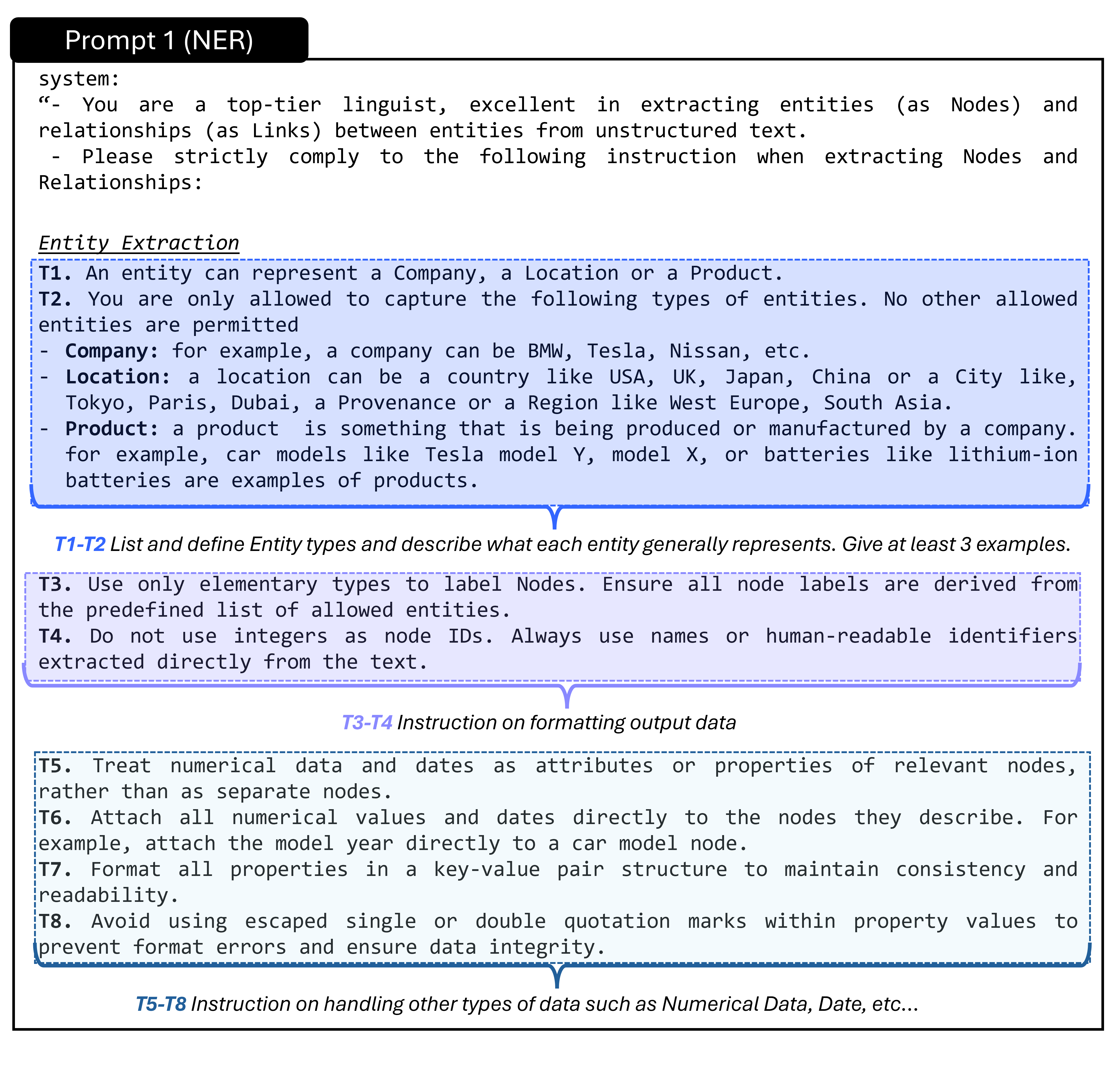}
    \caption{ Prompt Details (Prompt 1) for Guiding GPT-4 in NER Tasks}
    \label{ner-prompt1}
\end{figure}

\subsection{Knowledge Graph Construction Process}
\subsubsection{Named Entity Recognition}
 Our framework leverages these capabilities through a Zero-Shot Learning (ZSL) approach. ZSL is a machine learning paradigm that enables AI models to recognize and categorize entities or concepts not explicitly encountered during training \citet{Kojima2022LargeReasoners}. This approach is particularly valuable in our context, as it allows for flexibility in entity extraction without the need for task-specific fine-tuning.
To effectively guide the LLM in performing NER tasks using ZSL, we developed a carefully crafted prompt (Prompt 1) that provides contextual descriptions of the target entities. This prompt design is crucial, as it enables the model to recognize and extract entities similar to those described, even if they were not part of its training data.

Figure \ref{ner-prompt1} illustrates the detailed instructions used to guide the LLM in entity extraction. We divided the instruction into eight distinct tasks, labeled T1 through T8. 

\begin{itemize}
    \item \textit{Entity Definition (T1-T2)}: We begin by instructing the model on the specific categories of entities to be recognized. As shown in Figure \ref{ner-prompt1}, we provide clear definitions and at least three examples for each entity type. Note that since we provide only descriptions and not specific examples of input text with corresponding correct outputs, our framework employs Zero-Shot prompting rather than Few-Shot.

    \item \textit{Output Formatting (T3-T4)}: To ensure compatibility with subsequent processes in our framework, we provide explicit instructions on data output formatting. This includes specifying a key-value pair structure for entity properties, which enhances readability and consistency. We also include guidelines to avoid formatting errors.

    \item \textit{Data Handling Instructions (T5-T8)}: The LLM is given specific instructions on how to handle various types of data encountered in unstructured texts. While our primary focus is on a limited set of predefined entities, we also direct the model to attach any additional relevant information as properties to these entities. Furthermore, we provide guidelines for node labeling to ensure that the outputs are easily interpretable and well-organized.
\end{itemize}

This methodological approach allows us to leverage the advanced capabilities of GPT-4 in a zero-shot setting, enabling efficient and accurate NER across diverse texts without the need for task-specific training data.

\subsection{Relation Extraction}
\begin{figure}[htbp]
\centering
\includegraphics[width=0.9\linewidth]{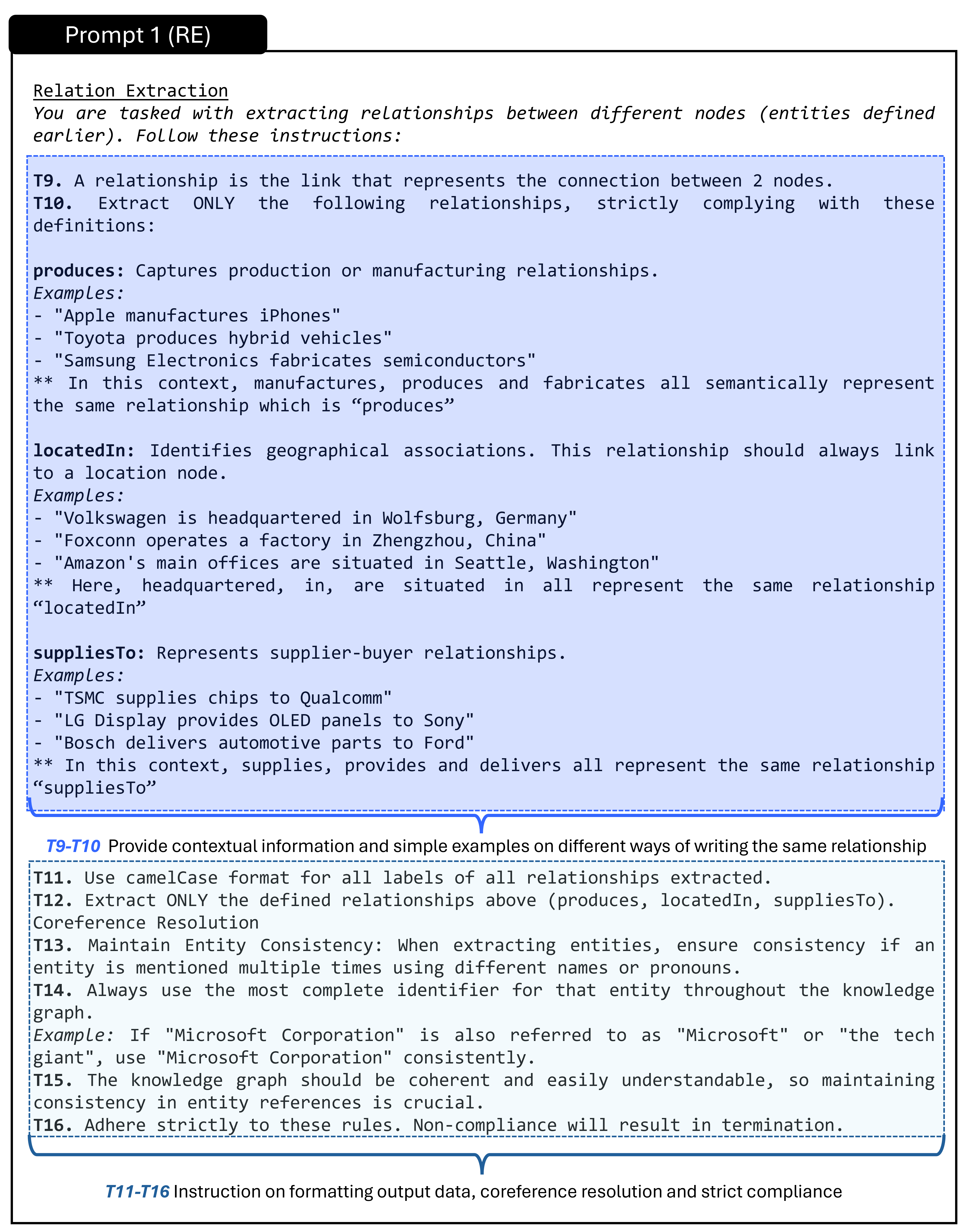}
\caption{ Prompt Details (Prompt 1) for Guiding GPT-4 in RE Tasks}
\label{re-prompt1}
\end{figure}

Similarly, for RE, we provide the LLM with specific instructions as shown in Figure \ref{re-prompt1}. 

We use the same prompt (Prompt 1) to continue listing the tasks for RE (T9-T16) as follows:

\begin{itemize}
    \item \textit{Task Definition (T9)}: Defines relationships as links between nodes. This basic definition helps the LLM understand the foundational concept of relationships in the context of our data.
    \item \textit{Targeted Relationships (T10)}: Specifies three types of relationships that the LLM must capture. For example, here, we list `Produces', `LocatedIn' and `SuppliesTo'. Each relationship is defined with description to ensure clarity. By providing semantic equivalents, we increase the model's ability to capture relevant variations of the same relationship.
    \item \textit{Label Formatting (T11-T16)}: Ensures consistency and readability in our data. This standardization is crucial for integrating the extracted data into our knowledge graph seamlessly.
\end{itemize}

\subsubsection{Entity Disambiguation}
Entity disambiguation is a critical process in ensuring the accuracy and integrity of the supply chain data within our knowledge graph. Given the diversity and inconsistency in naming conventions across various data sources, it is essential to correctly identify and merge entities that refer to the same underlying node. 

\begin{figure}[htbp]
    \centering
        \centering
        \includegraphics[width=0.9\linewidth]{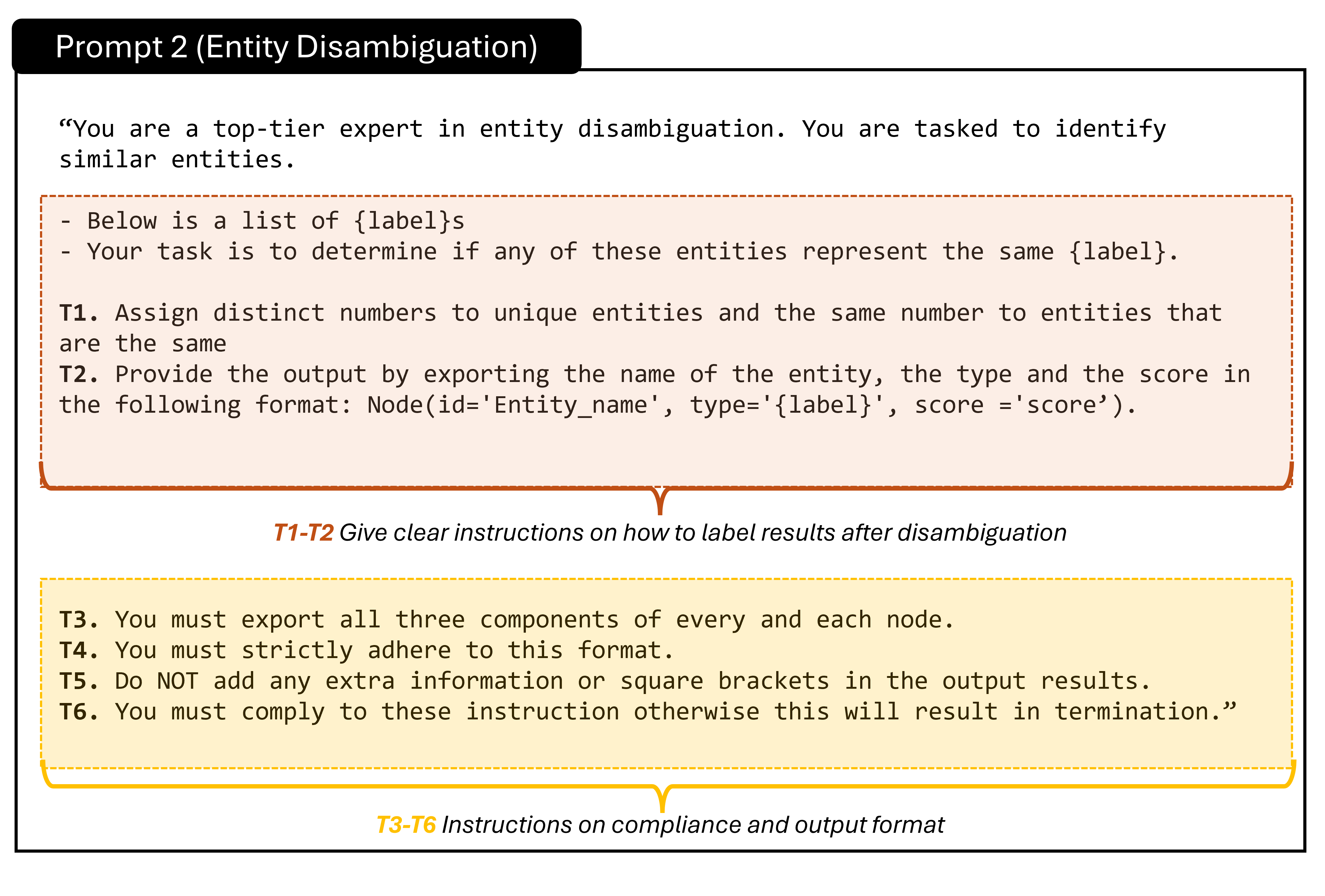}
        \caption{Customized Prompt (Prompt 2) for Guiding GPT-4 in Entity Disambiguation Tasks}
        \label{dis-prompt}
    \end{figure}

To effectively address this challenge, we have developed a detailed and structured prompt for our LLM, aimed at disambiguating entities identified in the initial extraction phase. This approach leverages the LLM's expertise in semantics and entity recognition to accurately unify entity representations (See Figure \ref{dis-prompt}).

\begin{figure}
        \centering
        \includegraphics[width=\linewidth]{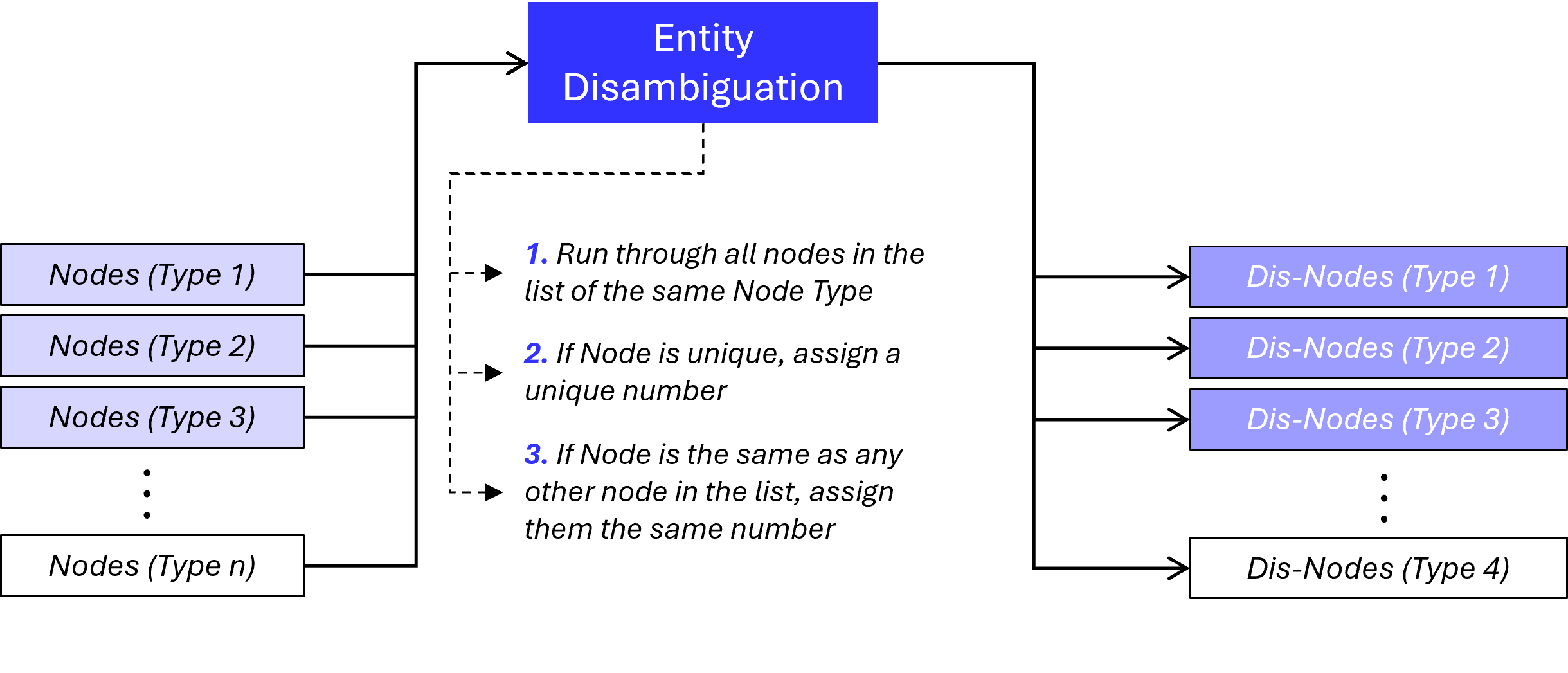}
        \caption{Flowchart of the Process for Entity Disambiguation}
        \label{dis-archit}
\end{figure}

The entity disambiguation process begins by instructing the LLM to assume the role of an expert in semantics and entity identification. As illustrated in Figure \ref{dis-archit}, we systematically categorize all nodes by type and present each node type list separately to the LLM for disambiguation. The LLM is then prompted to assign unique numerical identifiers to entities that represent distinct nodes, while allocating identical numbers to entities that semantically represent the same node. This approach leverages the LLM's contextual understanding to resolve naming inconsistencies and merge semantically equivalent entities.

\begin{figure}[htbp]
    \centering
    \includegraphics[width=1\linewidth]{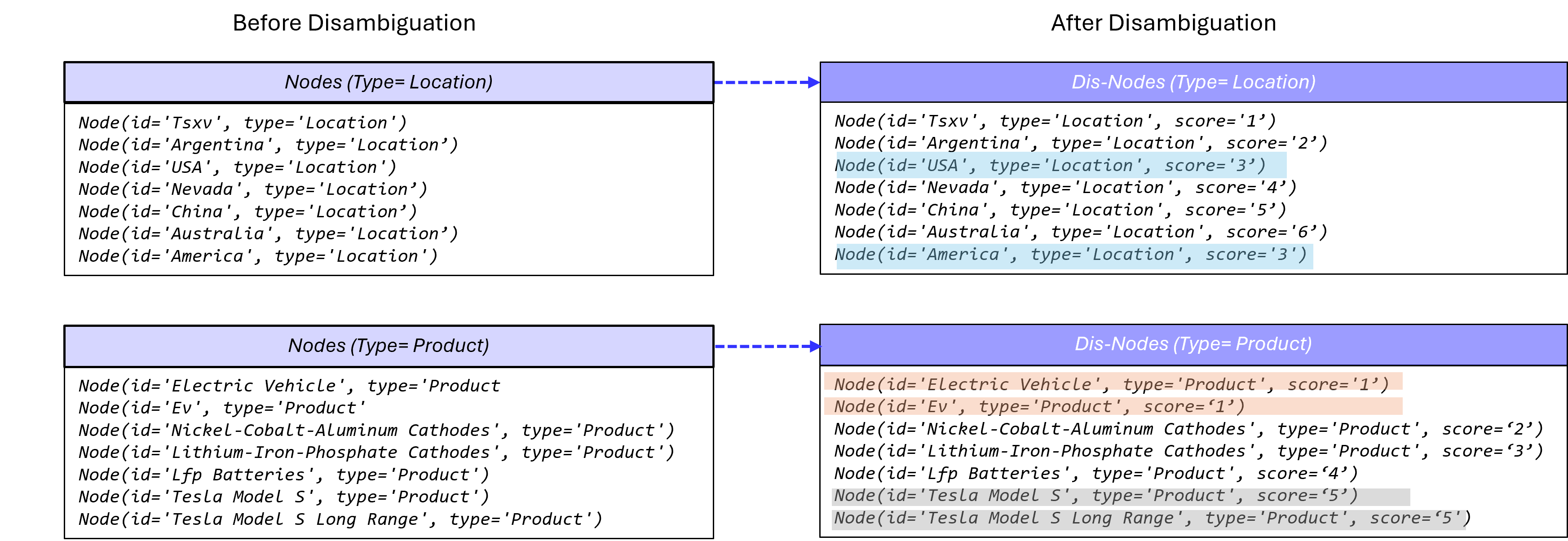}
    \caption{Results Example of Nodes Before and After Entity Disambiguation}
    \label{dis-example}
\end{figure}

Figure \ref{dis-example} shows some examples of the outcomes of the disambiguation process. We show how various entity types, such as locations and products, are distinctly identified and scored based on their semantic uniqueness. 
This method enables our framework to maintain a consistent and accurate representation of entities across diverse data sources, crucial for constructing a coherent and reliable supply chain knowledge graph.

\section {Experimental Studies}

In this section, we first describe the evaluation methods used to assess our proposed framework. We utilize accuracy as the primary evaluation metric and further analyze the framework’s consistency in generating reliable results. Following this, we present the quantitative results obtained from our framework. Additionally, we conduct a qualitative evaluation using a case study focused on the electric vehicle supply chain to demonstrate the framework’s effectiveness in enhancing supply chain visibility in domain-specific contexts. Specifically, we examine whether the framework can track critical minerals and mining companies involved in the production of electric vehicle batteries, thereby improving the visibility of Original Equipment Manufacturers (OEMs) into their extended supplier network.

\subsection{Evaluation Methods} 

In this section, we outline the evaluation methods used to assess our proposed framework's performance. We first describe the dataset developed for the evaluation and then detail the metrics employed to measure the framework's accuracy and consistency.

\subsubsection{Evaluation Dataset} 

Given the lack of benchmark datasets specifically designed for supply chain data, we developed a tailored evaluation methodology to assess the performance of our framework.

For our evaluation, we created a dataset of different Wikipedia pages relevant to electric vehicle manufacturers, comprising of a total of 1,277 sentences. We chose Wikipedia because it is a comprehensive, diverse repository of free-access information, which supports a broad view across industries, technologies, and supply chain relationships. This selection was specifically tailored to align with our subsequent proof of concept case study, which focuses on enhancing visibility in the electric vehicle supply chain, particularly in tracing the sourcing of raw materials for batteries from mines through tier-2 and tier-3 suppliers. Guided by the objectives of our case study, we strategically selected company profiles that span mining companies, battery manufacturers, and electric vehicle producers. This selection was made to maximize the likelihood of capturing comprehensive data on our predefined set of six entity types: Company, Location, Material, Person, Product, and Mine, as well as four key relationship types: locatedIn, suppliesTo, owns, and produces. 

It is important to clarify that in this evaluation, our main objective is to assess the LLM's ability to accurately extract entities and relationships, represented as triples, from textual content. The focus is on the technical extraction abilities of the LLM, not on the empirical accuracy of the information represented in the triples themselves. Researchers should carefully vet and select data to suit specific applications and contexts.

\begin{figure}[htbp]
    \centering  
    \subfigure[%
        \parbox{0.48\textwidth}{\centering Distribution of Node Types in the Validation Dataset}]{%
        \includegraphics[width=0.48\textwidth]{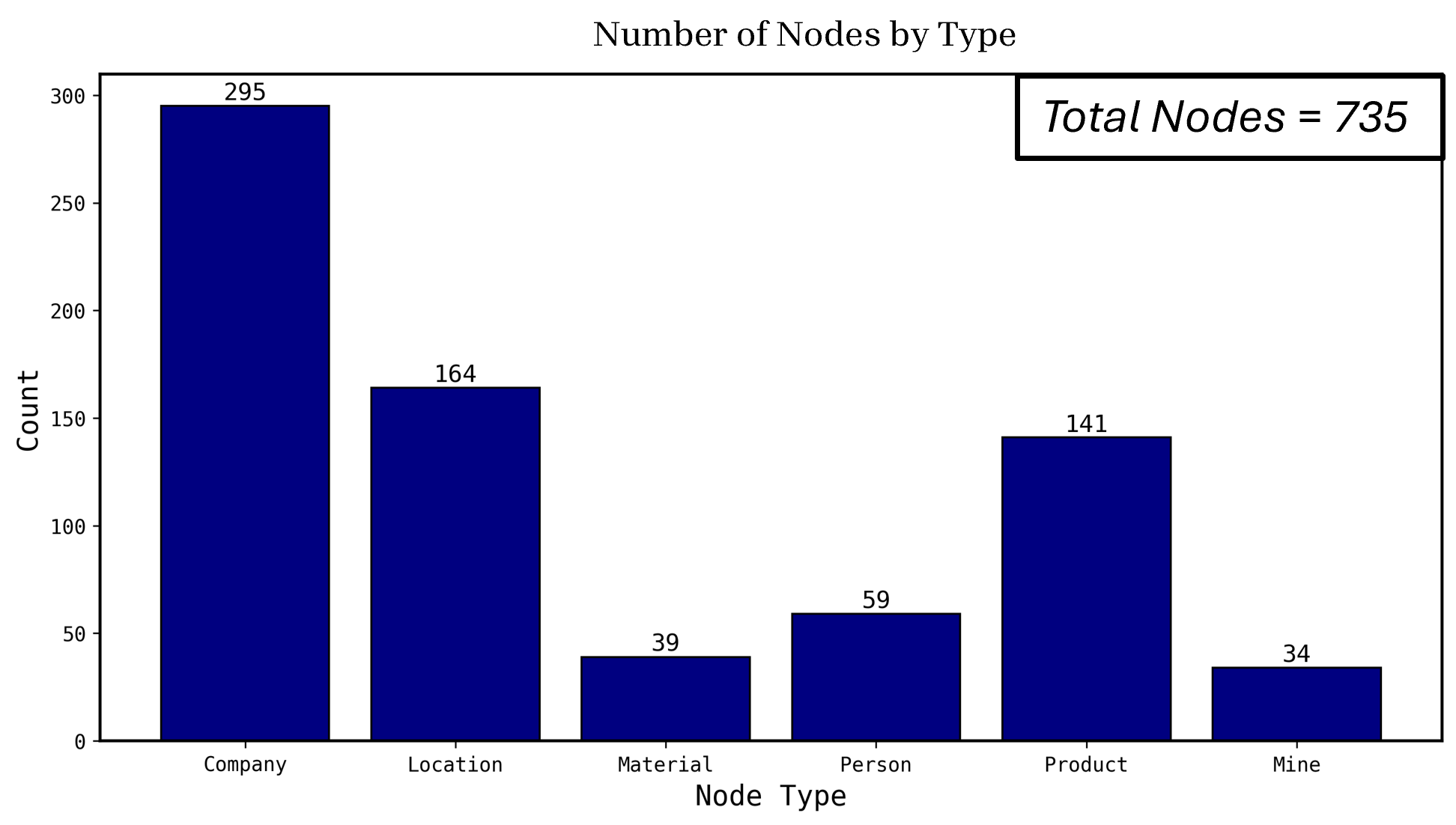}
        \label{fig:numberNodes}
    }
    \hfill  
    \subfigure[%
        \parbox{0.48\textwidth}{\centering Distribution of Relation Types in the Validation Dataset}]{%
        \includegraphics[width=0.48\textwidth]{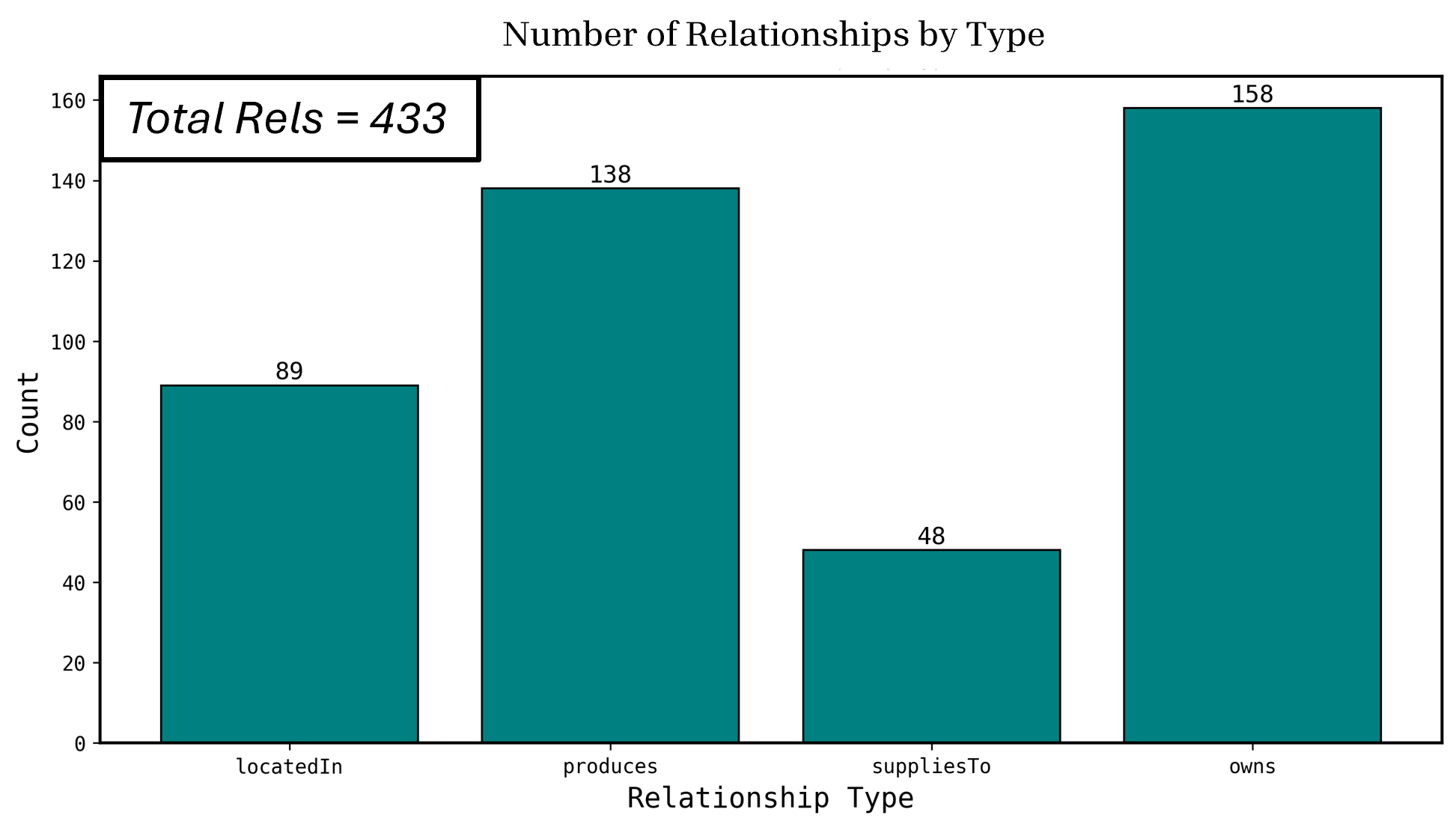}
        \label{fig:numberRels}
    }
    \caption{Validation Dataset Distributions}

\label{extraction_distribution}
\end{figure}

From the analyzed dataset, our framework successfully identified 735 nodes and 433 relationships, with their distribution detailed in Figure \ref{extraction_distribution}.

\subsubsection{Evaluation Metrics}

Our evaluation methodology is tailored to address the specific challenges and requirements of supply chain information extraction. We focus on assessing the accuracy of the extracted nodes and relationships, as these form the core of our knowledge graph and directly impact supply chain visibility and decision-making processes. We acknowledge that precision, recall and f-scores, which consider the missed triplets, are also important metrics for comprehensive evaluation. However, the manual identification of missed triplets is resource-intensive and currently not feasible due to the lack of benchmark datasets. Our current evaluation method serves as a preliminary step, demonstrating the framework's ability to extract valuable and accurate information. By focusing on accuracy and ensuring the correctness of the generated triplets, we provide a solid foundation for the practical applicability of our framework. This approach allows us to iteratively improve the system and gradually incorporate more comprehensive evaluation metrics as the field evolves. 

We measure accuracy across three key tasks as follows:

\textit{NER:}
\begin{itemize}
\item Correct: A node is correctly defined, and its type is accurate (e.g., identifying Tesla as a \textit{(Company)}).
\item Incorrect: A node is incorrectly classified (e.g., identifying Tesla as a \textit{(Location)} instead of a \textit{(Company)}).
\end{itemize}

Accuracy is calculated using:

\begin{equation}
\text{Accuracy}_{\text{NER}} = \frac{\text{Number of Correct Nodes Identified}}{\text{Total Nodes Identified (Correct + Incorrect)}}
\end{equation}

\textit{RE:}
\begin{itemize}
\item Correct: A relationship is accurately defined between two suitable nodes (e.g., Tesla \textit{(Company)} produces Model 3 \textit{(Product)}).
\item Incorrect: A relationship is incorrectly defined between two nodes, or the direction of the relationship is incorrect (e.g., identifying Tesla as supplying to CATL instead of CATL supplying to Tesla).
\end{itemize}

Accuracy is defined as:

\begin{equation}
\text{Accuracy}_{\text{RE}} = \frac{\text{Number of Correct Relationships Identified}}{\text{Total Relationships Identified (Correct + Incorrect)}}
\end{equation}

\textit{Entity Disambiguation:}
\begin{itemize}
\item Correct: One or more entities are correctly merged into a unique node, indicating they are similar, or an entity is correctly identified as unique with no similar nodes.
\item Incorrect: An entity is incorrectly merged with another node that does not represent the same unique entity, or it is not merged when it should have been.
\end{itemize}

Accuracy is determined by:
\begin{equation}
\text{Accuracy}_{\text{ED}} = \frac{\text{Number of Correctly Disambiguated Nodes}}{\text{Total Nodes Disambiguated (Correct + Incorrect)}}
\end{equation}

The consistency of our LLM-based framework is a critical factor in establishing its reliability for supply chain visibility applications. To assess this consistency, we conducted a comprehensive evaluation across two key dimensions: NER and RE. For each document in our dataset, we performed seven independent runs of the framework, allowing us to quantify variability and assess the robustness of our method across multiple iterations. 

Furthermore, we classified nodes by type (e.g., company, location, material, mine, product) and computed the standard deviation of counts for each unique node type. This granular analysis allows us to pinpoint any inconsistencies in specific entity type recognition. Similarly, for RE, we counted the total relationships identified in each run, calculated the standard deviation of these counts, and further categorized relationships by type (e.g., produces, suppliesTo, locatedIn). The standard deviation for each unique relationship type was also computed, providing insights into the consistency of specific relationship classifications.

\subsection{Results}

This section presents a comprehensive analysis of our proposed framework's performance, evaluated using accuracy metrics across three critical dimensions: NER, RE, and entity disambiguation. 

\subsubsection{Accuracy}

\begin{figure}[htbp]
    \centering  
    \subfigure[%
        \parbox{0.48\textwidth}{\centering Accuracy Metric for NER, RE, and Entity Disambiguation}]{
        \includegraphics[width=0.48\textwidth]{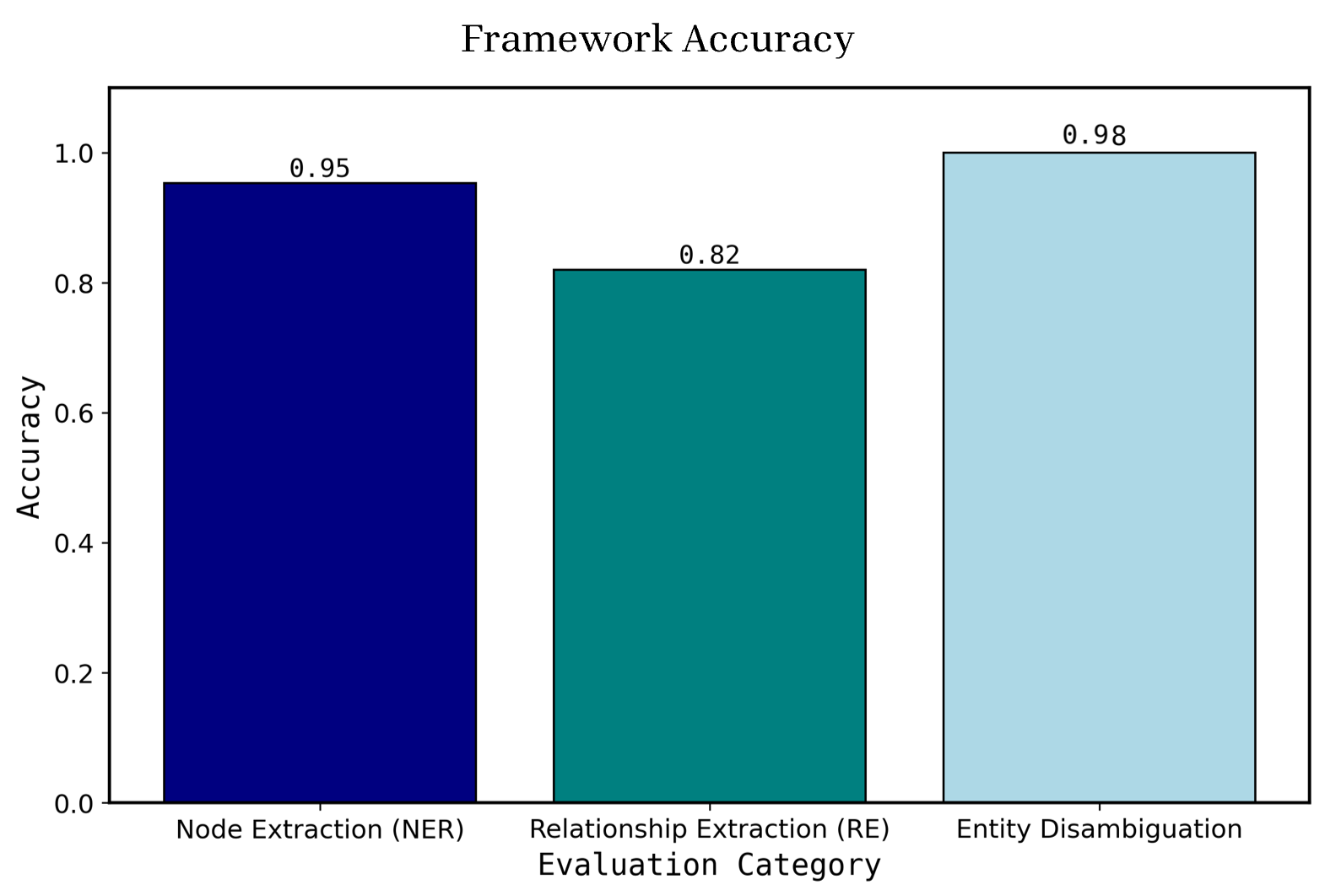}
        \label{fig:performanceMetrics}}
    \hfill  
    \subfigure[%
        \parbox{0.48\textwidth}{\centering Node Counts Before and After Entity Disambiguation}]{%
        \includegraphics[width=0.48\textwidth]{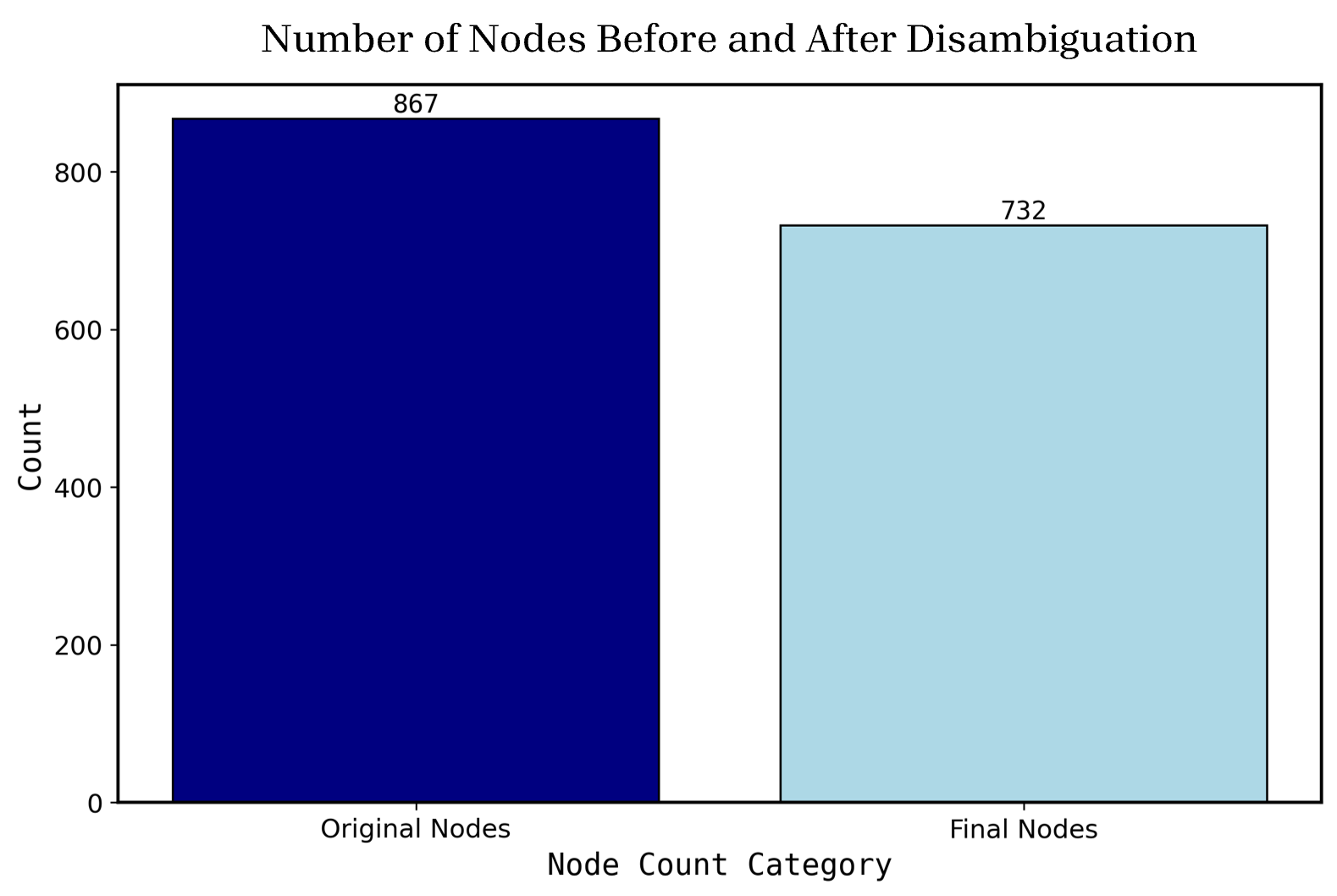}
        \label{fig:nodeCounts}
    }
    \caption{Overall Performance and Node Counts Before and After Disambiguation}
    \label{overallPerformance}
\end{figure}

Figure \ref{overallPerformance} shows the performance of our framework using Zero-Shot Learning across the three key tasks defined.

NER achieved a high accuracy of 0.95, demonstrating the LLM's robust capability in identifying diverse entity types within unstructured text. This performance can be attributed to the LLM's extensive pre-training on diverse datasets, coupled with our carefully crafted prompts that provide clear supply chain entity definitions and examples.

RE, while achieving the lowest accuracy among the three tasks at 0.82, it still demonstrates significant capability in a zero-shot setting. This performance is noteworthy given the complexity of identifying semantic relationships without task-specific training. The relatively lower accuracy in RE compared to NER suggests that relationship identification is a more challenging task, potentially due to the nuanced ways relationships can be expressed in natural language. Future work could explore enhancing RE performance through few-shot learning or by providing a set of diverse examples in the prompts.

Entity Disambiguation recorded the highest accuracy at 0.98, showcasing the LLM's exceptional ability to recognize semantic similarities and differences among entities. The framework successfully reduced the initial 867 nodes to 732 unique entities, merging 135 duplicates. This high performance in disambiguation is crucial for maintaining data integrity and reducing redundancy in the knowledge graph, a key factor in enhancing supply chain visibility.

\begin{figure}[htbp]
    \centering  
    \subfigure[%
        \parbox{0.48\textwidth}{\centering NER Accuracy by Node Type}]{%
        \includegraphics[width=0.48\textwidth]{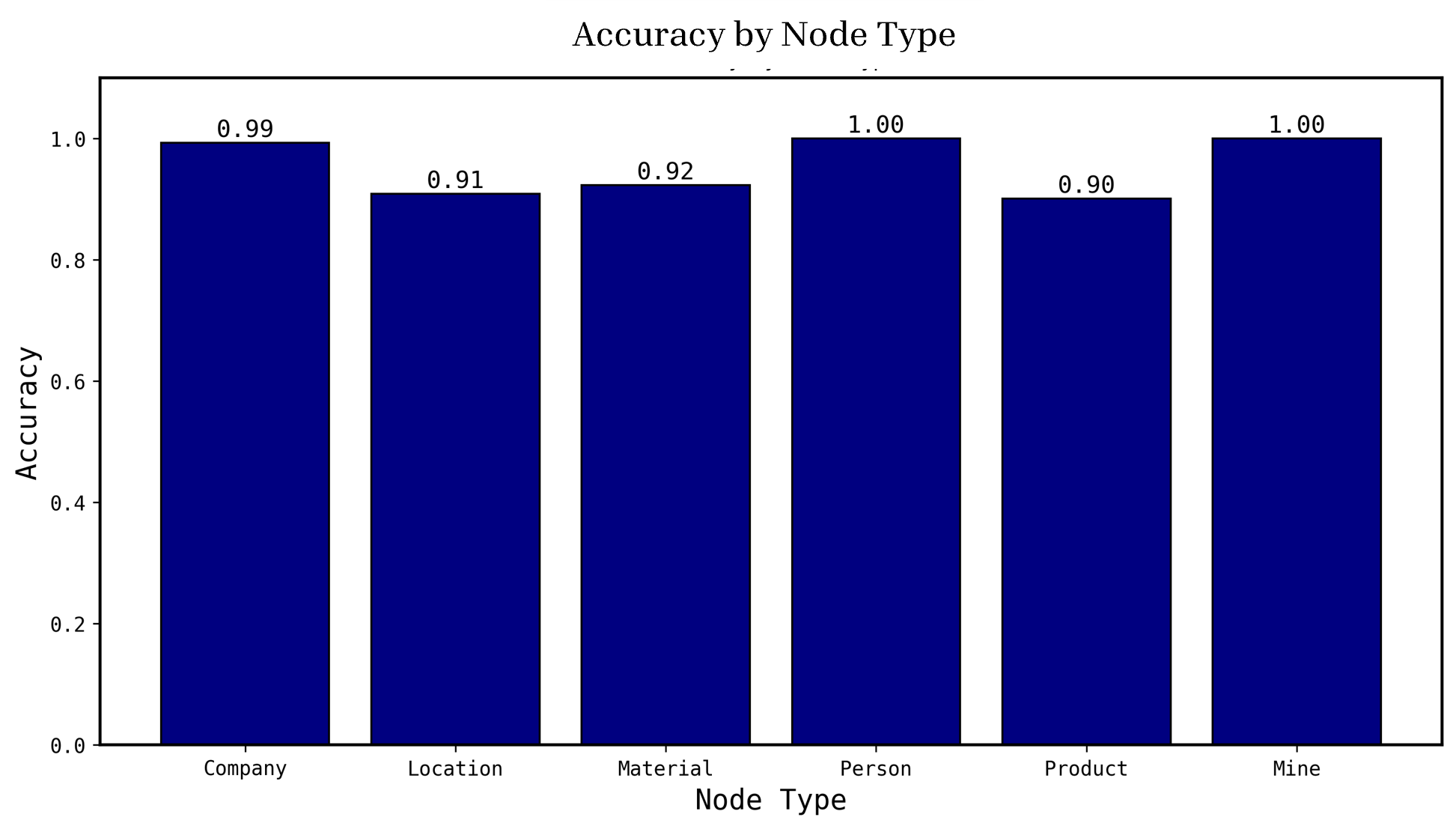}
        \label{accuracy-by-type-a}
    }
    \hfill  
    \subfigure[
        \parbox{0.48\textwidth}{\centering RE Accuracy by Relation Type}]{%
        \includegraphics[width=0.48\textwidth]{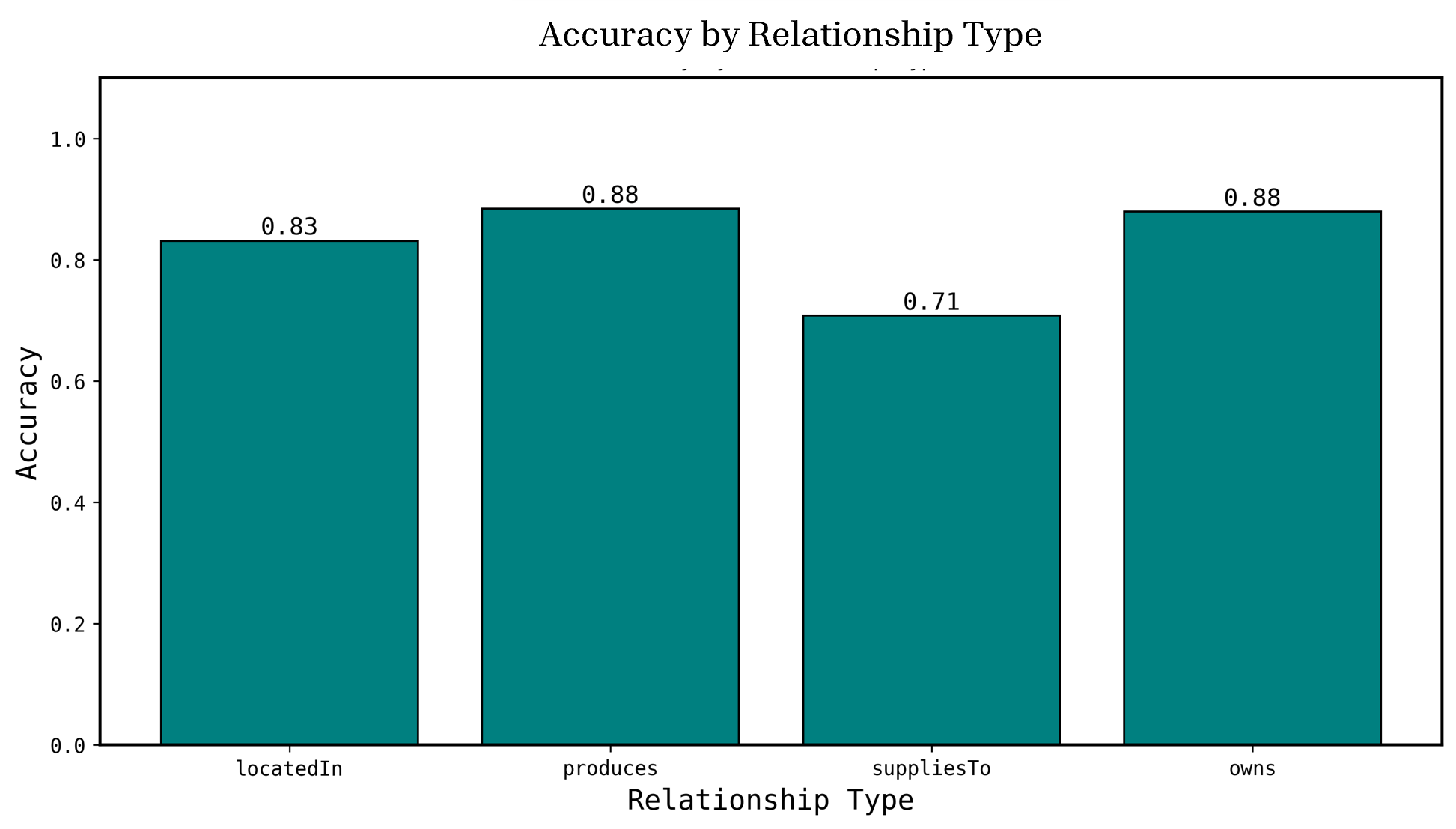}
        \label{accuracy-by-type-b}
    }
    \caption{Accuracy of NER and RE by Node and Relation Type}
    \label{accuracy-by-type}
\end{figure}

We further analyze the accuracy achieved by node and relationship types to understand the LLM's performance in capturing different entities and relationships within the supply chain. As shown in Figure \ref{accuracy-by-type-a}, the LLM demonstrates high accuracy in capturing various node types. Particurarly in identifying nodes such as companies, products, and mines, achieving near-perfect accuracy. This performance can be attributed to the LLM's extensive pre-training on diverse datasets, which equips it with a robust understanding of common entity types and their contextual usage in text \citet{Kojima2022LargeReasoners, Brown2020LanguageLearners}. 

Furthermore, for RE, we observe that the lowest accuracy is recorded for the `suppliesTo' relationship, with an accuracy of 0.71 \ref{accuracy-by-type-b}. This lower performance can be attributed to the diverse ways in which supply chain relationships are expressed in natural language. The `suppliesTo' relationship, in particular, can be represented using various terminologies and syntactic structures, making it challenging for the LLM to consistently identify and extract this relationship.

Despite prompting the LLM to capture different semantic equivalents of the `suppliesTo' relationship, the large inherent variability in language used in supply chain poses a challenge. To address this, future work could explore the use of few-shot learning techniques, where the LLM is provided with explicit examples to cover the majority of different ways this relationship can be represented in the supply chain context. This approach has been shown to improve the performance of LLMs in similar tasks by providing more targeted training data \cite{Brown2020LanguageLearners, OpenAI2023GPT-4Report}.
Additionally, defining and constraining the types of nodes that the `suppliesTo' relationship should link to could further enhance the model's accuracy. By providing more specific guidelines on the expected relationships between entities, the LLM can better focus its extraction efforts and reduce the likelihood of errors.

The framework's performance across all three tasks, particularly in a zero-shot setting, demonstrates the potential of LLMs in extracting structured information from unstructured text when guided by effective prompting. This approach offers significant flexibility, allowing for easy adaptation to new entity types or relationships without the need for retraining.

\subsubsection{Consistency}
\begin{figure}[htbp]
    \centering  
    \subfigure[%
        \parbox{0.48\textwidth}{\centering NER Consistency Across 7 Runs}]{%
        \includegraphics[width=0.48\textwidth]{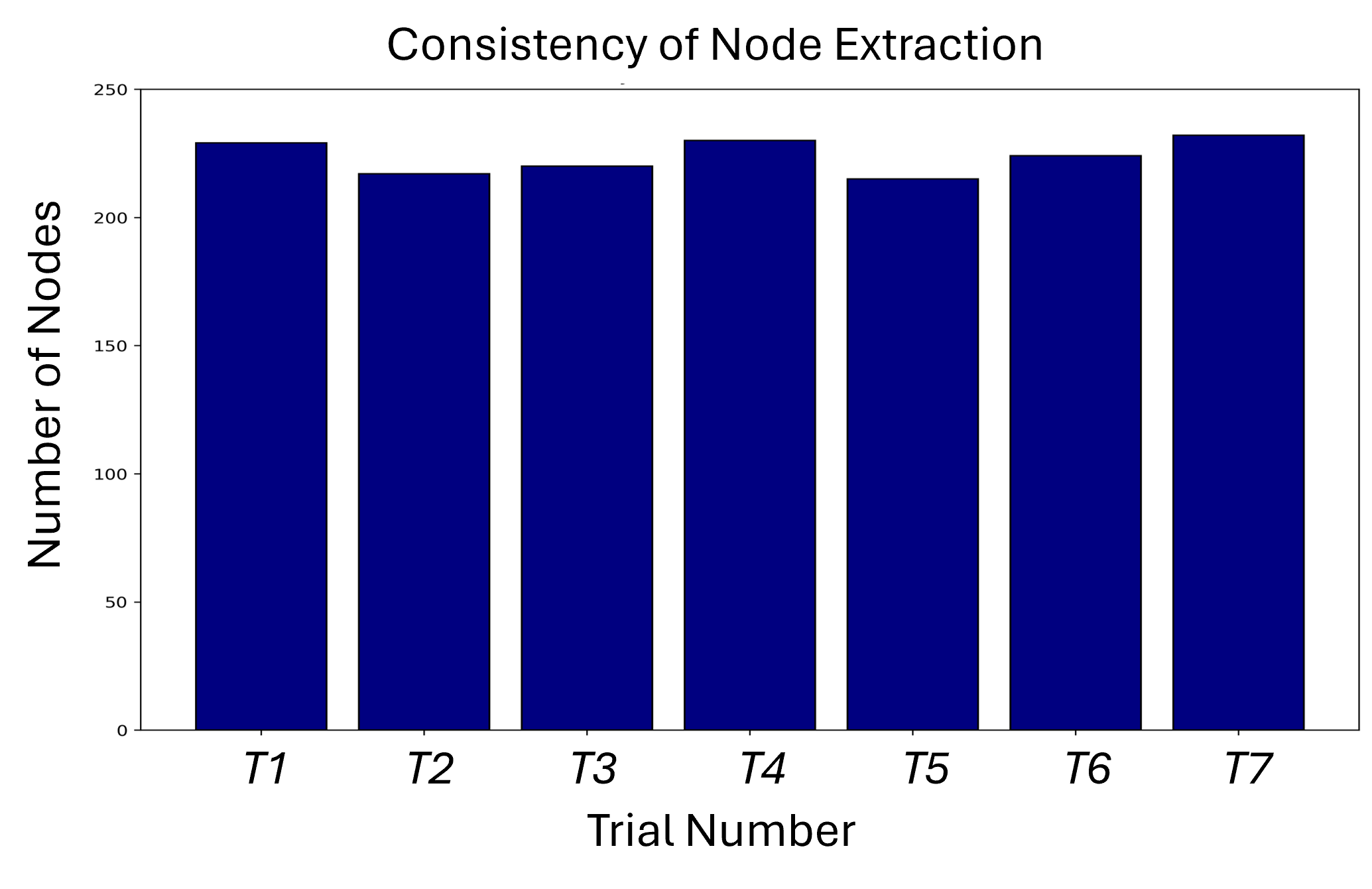}
        \label{fig:nodeExtractionVariability}
    }
    \hfill  
    \subfigure[
        \parbox{0.48\textwidth}{\centering RE Consistency Across 7 Runs}]{%
        \includegraphics[width=0.48\textwidth]{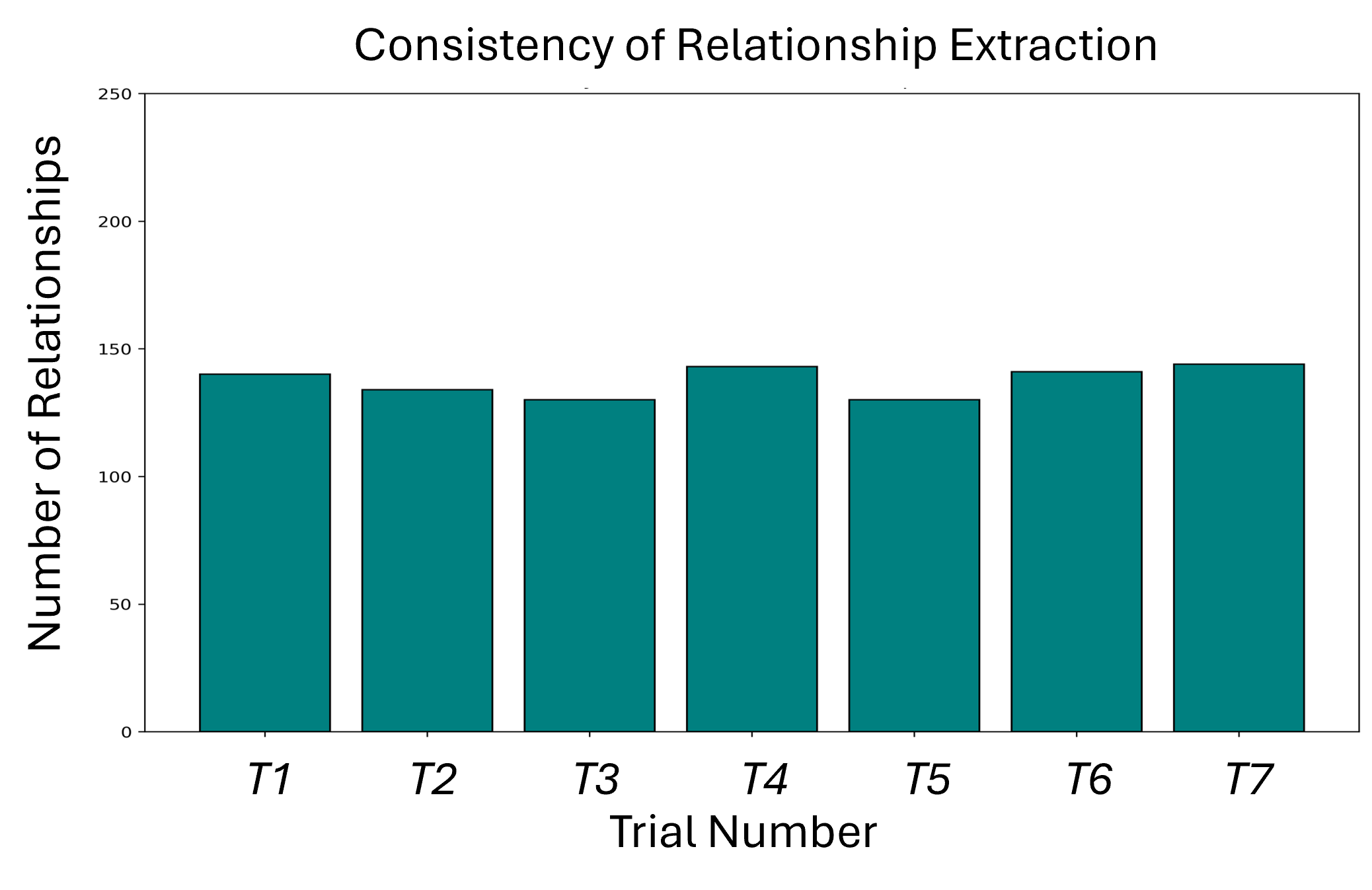}
        \label{fig:nodeExtractionVariability}
    }
    \caption{Variability in Node and Relationship Extraction Across Seven Framework Runs on a Fixed Dataset}
    \label{consistency-performance}
\end{figure}

Figure \ref{consistency-performance} shows the counts of nodes and relationships extracted over 7 distinct runs. We observe low fluctuations between different trials for both node and RE tasks.

To further quantify the consistency of our framework, we calculate the Mean, Standard Deviation, Coefficient of Variation and Range for both tasks across the 7 different trials as shown in Table \ref{Table-1}. As observed, the low coefficients of variation for both node count (0.03) and relationship count (0.04) indicate high consistency in the framework's performance. These values suggest that the variability in extraction is minimal relative to the mean, demonstrating the framework's reliability in producing consistent results.
The standard deviations of 6.22 and 5.55 for node and relationship counts, respectively, compared to their respective means of 223.86 and 137.43, further support the framework's consistency. The relatively small ranges (17 for nodes and 14 for relationships) also indicate that the framework's performance remains stable across different runs.

\begin{table}[htbp]
\centering
\begin{threeparttable}
\caption{Statistical metrics to quantify the consistency of the framework in NER and RE}
\label{Table-1}
\begin{tabular}{lcc}
\toprule
Metric & Node Count & Relationship Count \\
\midrule
Mean & 223.86 & 137.43 \\
Standard Deviation & 6.22 & 5.55 \\
Coefficient of Variation & 0.03 & 0.04 \\
Range & 17.00 & 14 \\
\bottomrule
\end{tabular}
\end{threeparttable}
\end{table}

We further break down the consistency metrics by node type. As shown in Table \ref{Table2}, Location nodes demonstrate the lowest CV of 0.438, among all types, indicating high consistency in identifying geographical entities.
Material nodes exhibit the highest CV of 0.960, which may be attributed to the diverse ways materials can be mentioned in text, potentially leading to more variable extraction.

\begin{table}[htbp]
\centering
\begin{threeparttable}
\caption{Statistical Analysis of Framework Consistency in NER by Type}
\label{Table2}
\begin{tabular}{@{}lcccc@{}}
\toprule
Node Type & Mean & Standard Deviation & Coefficient of Variation & Range \\ \midrule
Company   & 16.83 & 10.41 & 0.618 & 35 \\
Location  & 7.37  & 3.23  & 0.438 & 10 \\
Person    & 7.49  & 4.85  & 0.648 & 14 \\
Product   & 8.20  & 6.57  & 0.801 & 21 \\
Material  & 5.56  & 5.34  & 0.960 & 12 \\
Mine      & 8.20  & 4.79  & 0.584 & 11 \\ \bottomrule
\end{tabular}
\end{threeparttable}
\end{table}

Table \ref{Table3} presents consistency metrics for relationship types. As we observe from the table, the relationship `locatedIn' shows the lowest CV (0.341), indicating high consistency in identifying geographical associations. On the other hand `owns' relationships have reported the highest mean (10.43) but also the highest CV (0.891), suggesting that ownership relationships are frequently identified but with considerable variability across trials. These results demonstrate that our LLM-based framework exhibits strong overall consistency in both node and relationship extraction.

\begin{table}[htbp]
\centering
\begin{threeparttable}
\caption{Statistical Analysis of Framework Consistency in RE by Type}
\label{Table3}
\begin{tabular}{@{}lcccc@{}}
\toprule
Relationship Type & Mean & Standard Deviation & Coefficient of Variation & Range \\ \midrule
produces     & 9.03 & 6.22 & 0.689 & 21 \\
locatedIn    & 3.46 & 1.18 & 0.341 & 4  \\
owns         & 10.43 & 9.29 & 0.891 & 29 \\
suppliesTo   & 3.97 & 2.44 & 0.616 & 10 \\ \bottomrule
\end{tabular}
\end{threeparttable}
\end{table}

The variability observed across different node and relationship types provides insights into areas where the framework excels (e.g., location identification) and areas that may benefit from further refinement (e.g., material identification and ownership relationships). These findings can guide future improvements to the framework, potentially through more targeted prompting. 

The high consistency demonstrated by our framework, especially in a zero-shot learning context, underscores the high potential of LLM-based approaches for supply chain visibility applications. It suggests that with carefully designed prompts, LLMs can reliably extract structured information from unstructured text, even without task-specific fine-tuning. This capability offers significant advantages in terms of flexibility and adaptability to diverse supply chain scenarios.

\subsection{Case-Study}

To demonstrate the practical benefits and effectiveness of our framework, we applied it to a focused case study aimed at enhancing the supply chain visibility of electric vehicle manufacturers. These supply chains are inherently complex, often extending beyond immediate tier-1 and tier-2 suppliers. Critical minerals like lithium, nickel, and cobalt, essential for electric vehicle batteries, are sourced from various global mining operations. However, the exact origins of these minerals are often unclear, posing significant challenges in times of geopolitical tensions or potential supply chain disruptions.
The primary goal of our case study was to evaluate whether our framework could effectively map out dependencies and generate actionable insights about the supply chain of critical minerals. To achieve this, we selected a representative group of companies from three key categories: electric vehicle manufacturers, battery suppliers, and mining companies. Our selection criteria were based on the prominence of these companies in their respective fields and their significance in the global supply chain for electric vehicle batteries. The selected companies included:
\begin{itemize}
    \item Electric Vehicle Manufacturers: Toyota, Tesla Inc., BMW, BYD Co., Ltd.
    \item Battery Suppliers: Samsung SDI, CATL, Panasonic
    \item Mining Companies: Zijin Mining, Albemarle Corporation, Tianqi Lithium, Ganfeng Lithium Co Ltd, Norilsk Nickel, Jinchuan Group
\end{itemize}

As this case study aims to provide a proof of concept for the proposed framework, we fetched information for these companies using Wikipedia pages. We prompted the LLM to capture various types of nodes (Company, Person, Location, Material, Mine, Product) and relationships (suppliesTo, contains, produces, locatedIn, owns). This data was then processed through our framework to map out the intricate web of supply chain relationships and dependencies. Se have focused on selected zoomed-in sections of supply networks to demonstrate the capabilities of our framework. We highlight the benefits and interdependencies that were successfully extracted and visualized. Representative examples from each category are detailed, with further cases available in the Appendices Section.

\subsubsection{Enhanced Visibility Beyond Tier-1 Suppliers}
Our framework can extend the visibility of many OEMs beyond their immediate tier-1 suppliers, uncovering the origins and mines of critical minerals used in electric vehicle manufacturing.

\begin{figure}[htbp]
    \centering  
    \subfigure[%
        \parbox{0.40\textwidth}{\centering General Illustration on Limited Extended Network Visibility.}]{%
        \includegraphics[width=0.48\textwidth]{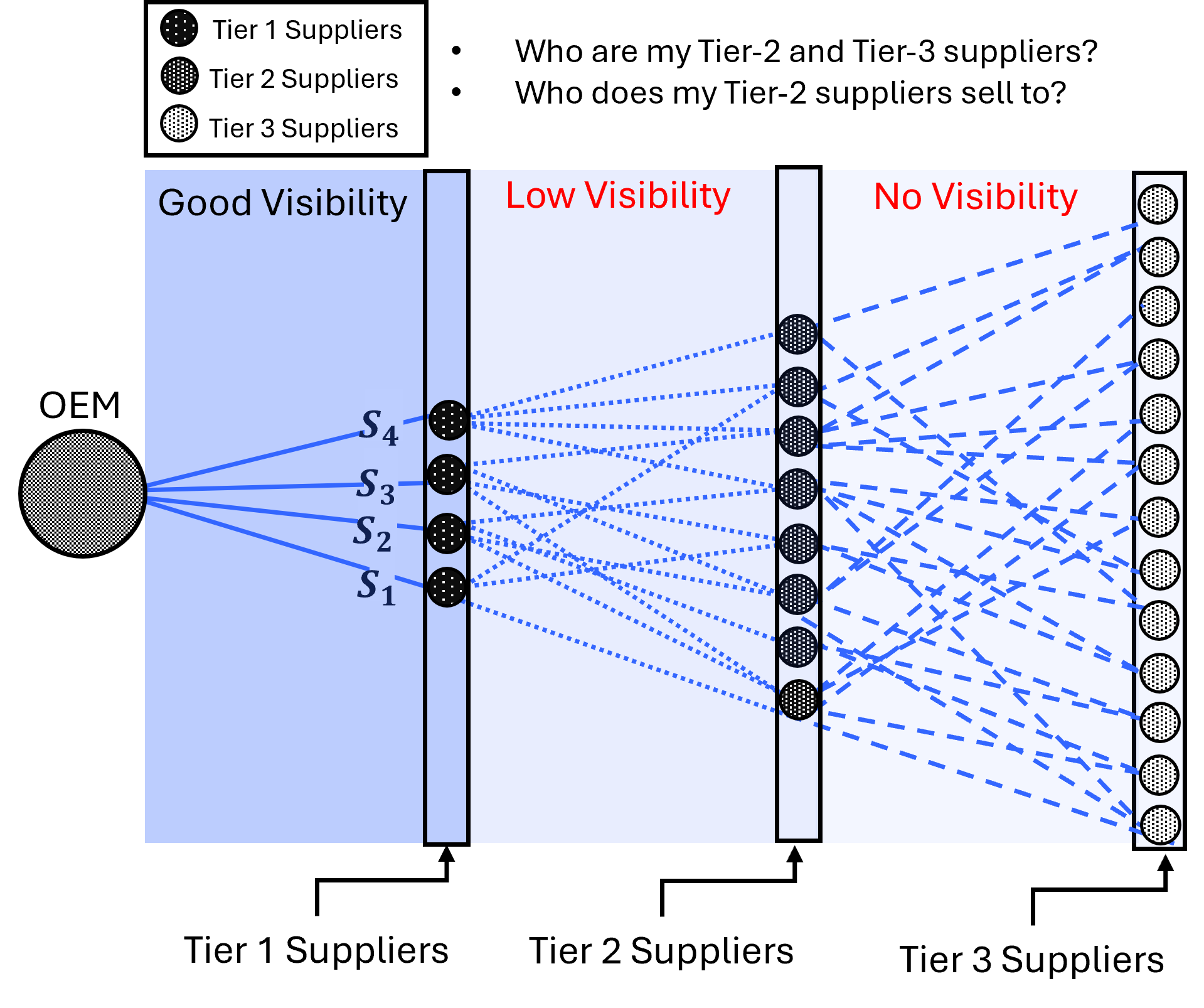}
        \label{fig:nodeExtractionVariability}
    }
    \hfill  
    \subfigure[
        \parbox{0.43\textwidth}{\centering Case-study Example: Capturing Beyond Tier-1 Visibility}]{%
        \includegraphics[width=0.48\textwidth]{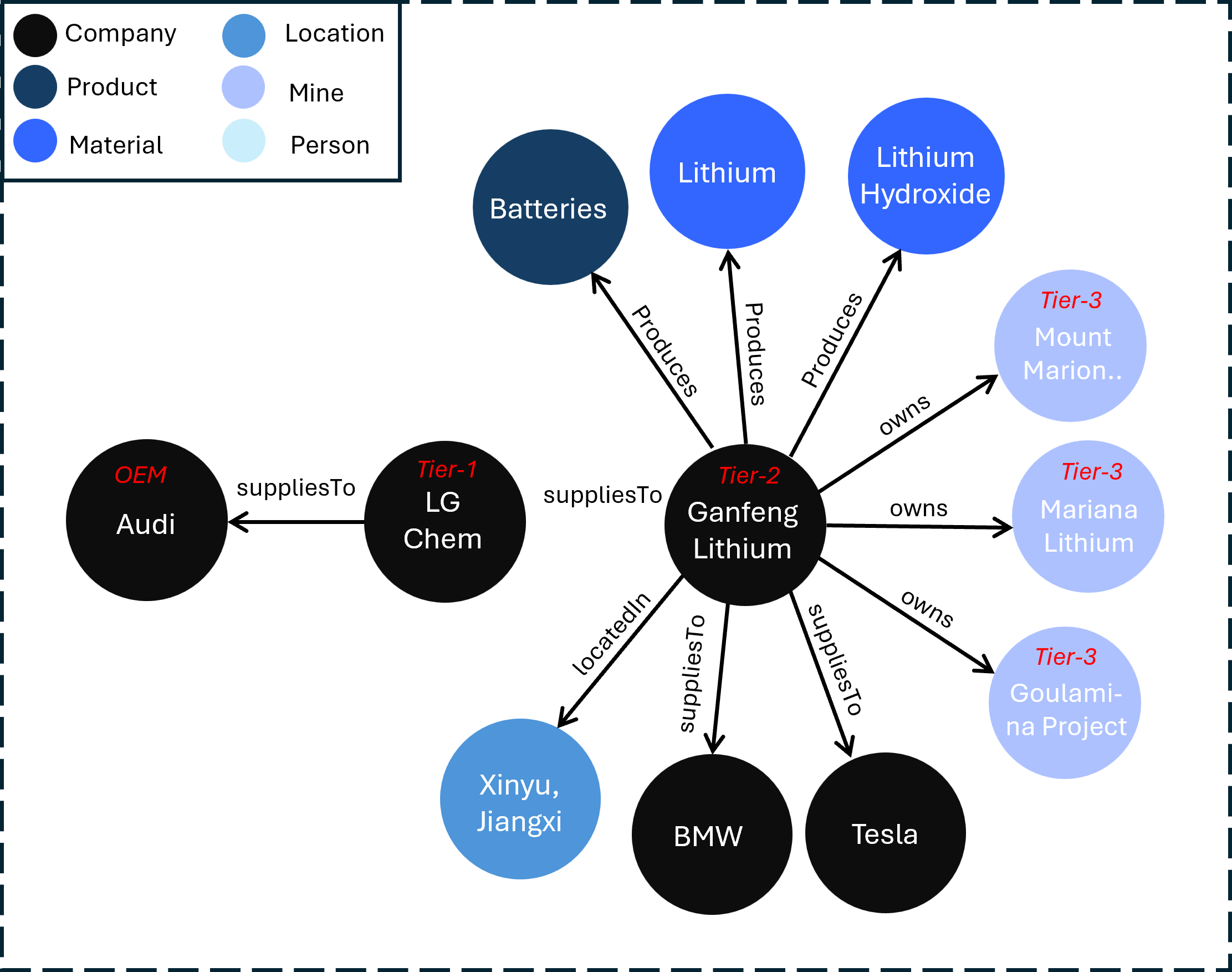}
        \label{fig:nodeExtractionVariability}
    }
    \caption{Enhanced Visibility Beyond Tier-1 Suppliers}
    \label{benifit1}
\end{figure}

As shown in Figure \ref{benifit1}, the framework captured Audi's tier-2 and tier-3 suppliers, revealing the specific lithium mines that supply the lithium used in Audi's electric batteries. For instance, Ganfeng Lithium, located in China, owns three different lithium mines. Additionally, the framework identified that Ganfeng Lithium also supplies to competitors of Audi, such as BMW and Tesla. \appendixautorefname{A} provides a more detailed network illustration and additional examples of extended tier visibility. Neo4j graph database \citet{Neo4j2024Neo4j:Platform} was used to visualize these dependencies. 

By tracing the supply chain of lithium, an OEM can identify the specific mines where their suppliers source lithium. This knowledge enables proactive measures to secure alternative sources if disruptions occur at the primary mine. This extended visibility allows OEMs to understand the full scope of their supply chains, from raw material extraction at mines to the production of battery components. Such insights are crucial for strategic planning and risk management, especially in scenarios involving geopolitical tensions or supply disruptions.

\subsubsection{Alternative Sourcing and Supplier Diversification}

\begin{figure}[htbp]
    \centering  
    \subfigure[%
        \parbox{0.40\textwidth}{\centering General Illustration on Limited on Alternative Suppliers}]{%
        \includegraphics[width=0.48\textwidth]{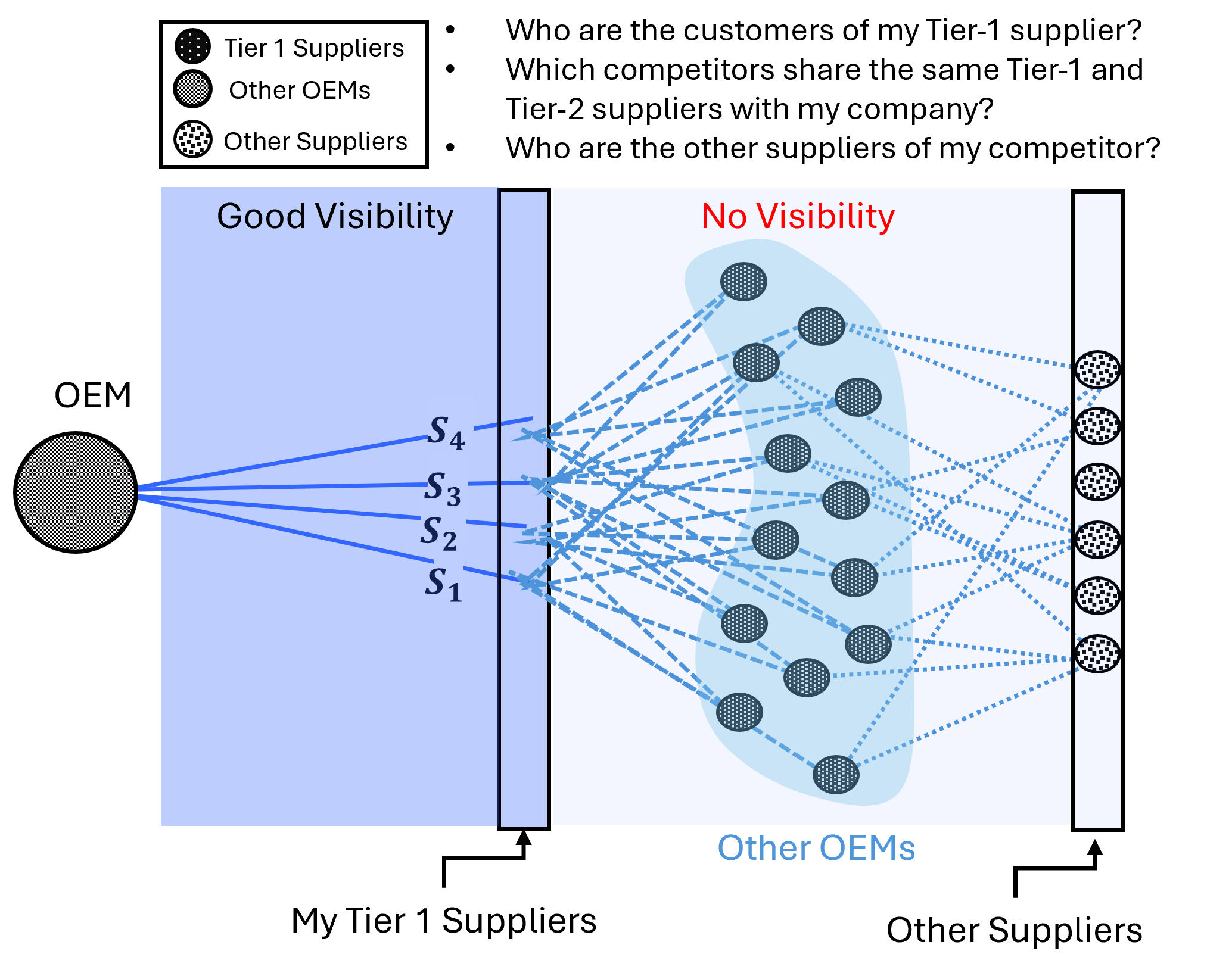}
        \label{fig:nodeExtractionVariability}
    }
    \hfill  
    \subfigure[
        \parbox{0.40\textwidth}{\centering Case-study Example: Capturing Shared and Other Alternative Suppliers in the Network}]{%
        \includegraphics[width=0.48\textwidth]{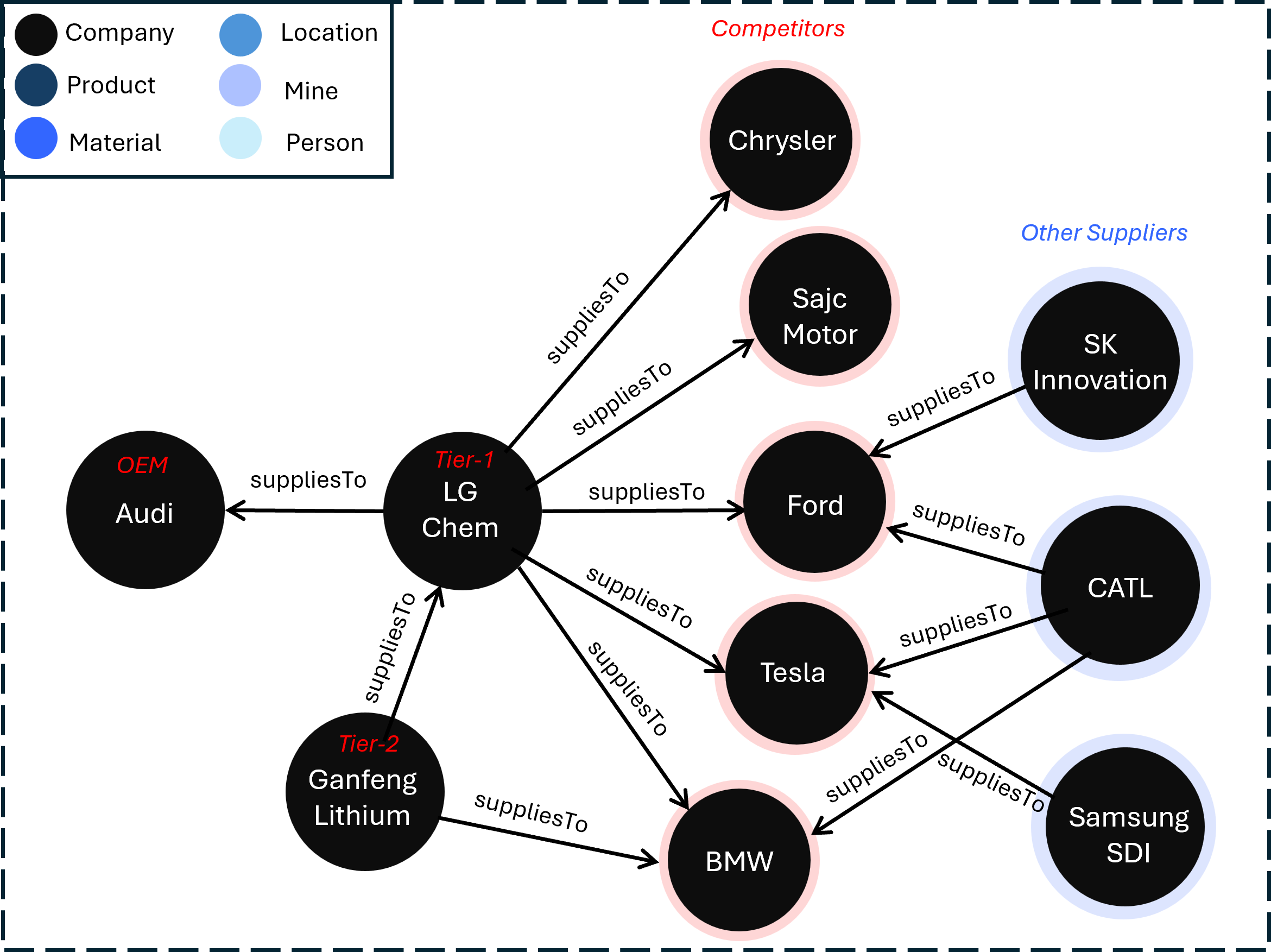}
        \label{fig:nodeExtractionVariability}
    }
    \caption{Alternative Sourcing and Supplier Diversification}
    \label{benifit2}
\end{figure}

As shown in Figure \ref{benifit2}, the framework also enables OEMs to visualize supply chain relationships and answer strategic questions such as "What other OEMs do my tier-1 suppliers supply to?" and "What alternative suppliers supply to my competitors?". As shown, LG-Chemical is Audi's tier-1 supplier but also supplies batteries to competitors like Chrysler, SAIC Motor, and Ford. Additionally, although Audi shares the same tier-2 supplier with Tesla, Ford, and BMW, these companies also have other suppliers such as CATL. \appendixautorefname{B} provides a more detailed network illustration and similar examples of visualizing alternative suppliers and shared suppliers with OEMs.

This could provide decision makers with valuable insights for supply chain optimization and strategic decision-making. By identifying that a key battery supplier also supplies to a competitor, an OEM can negotiate better terms or seek alternative suppliers to gain a competitive edge.

\subsubsection{Mapping of Product Supply Chain Network}

We present another application of the framework to demonstrate its capability to capture where certain materials are located, who owns and distributes these materials, and which mines they come from.

\begin{figure}[htbp]
    \centering  
    \subfigure[%
        \parbox{0.40\textwidth}{\centering General Illustration on Mapping Supply Chain Network of a Product or Material}]{%
        \includegraphics[width=0.48\textwidth]{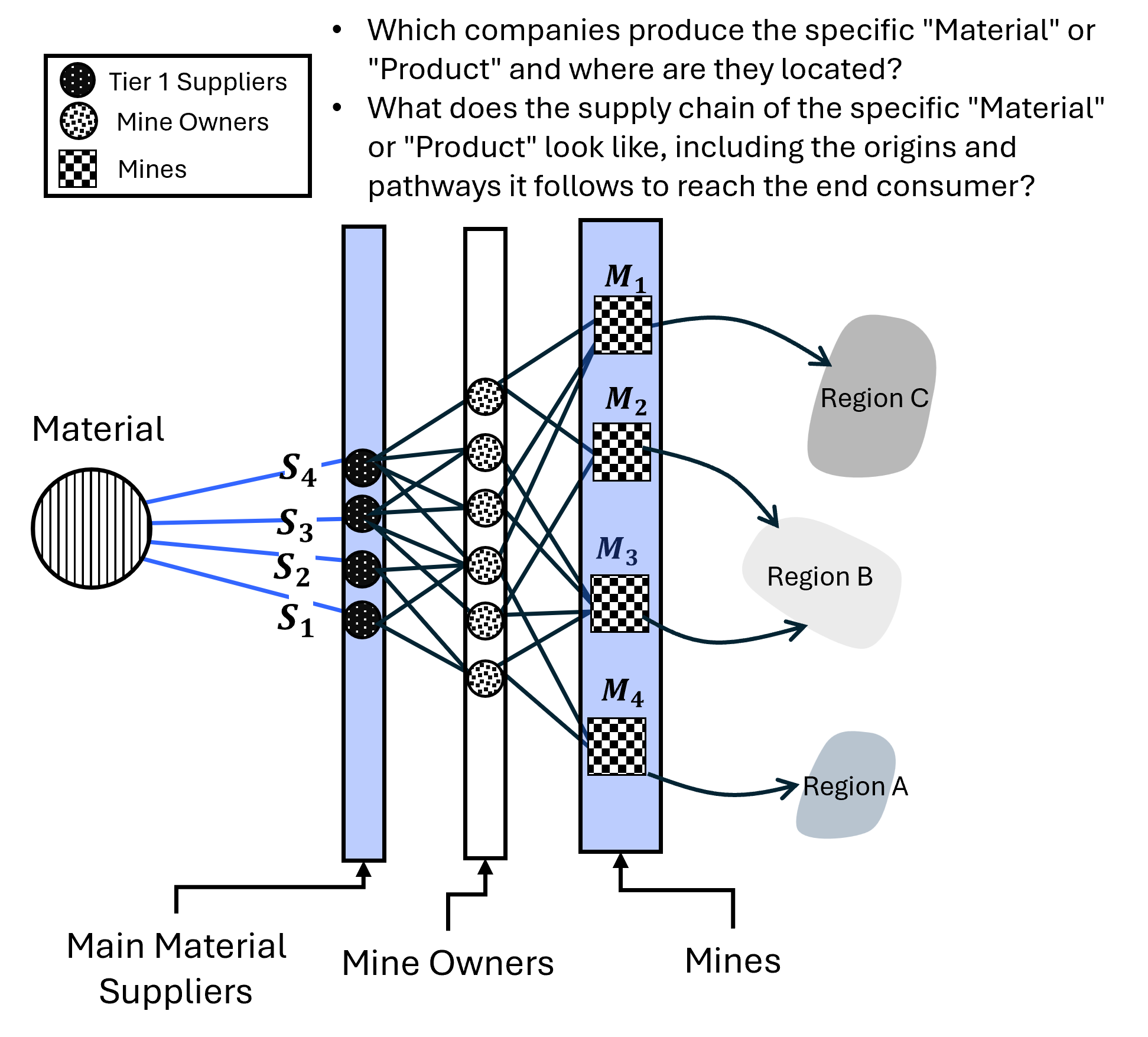}
        \label{fig:nodeExtractionVariability}
    }
    \hfill  
    \subfigure[
        \parbox{0.40\textwidth}{\centering Case-study Example: Capturing Nickel's Supply Chain Network}]{%
        \includegraphics[width=0.48\textwidth]{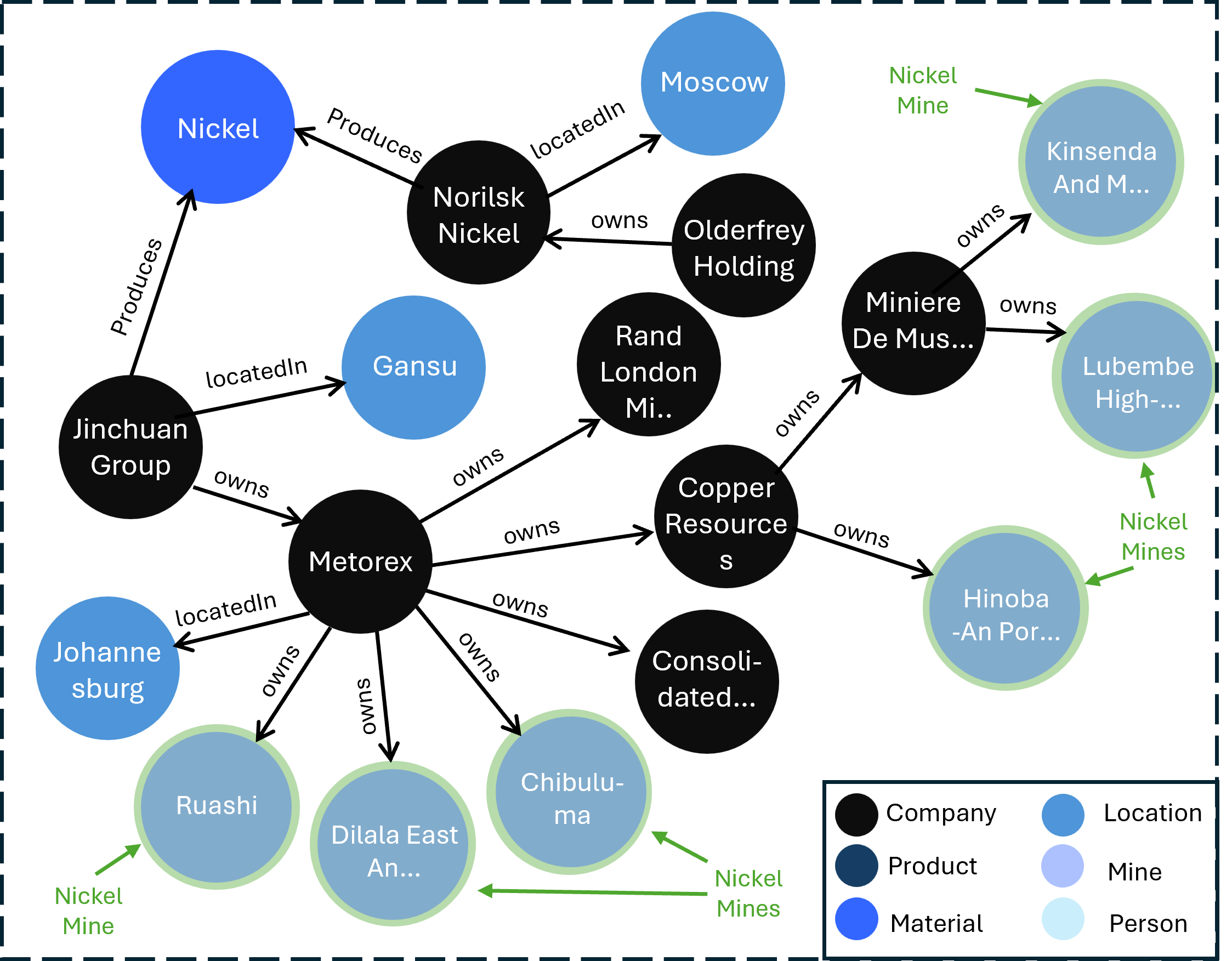}
        \label{fig:nodeExtractionVariability}
    }
    \caption{Alternative Sourcing and Supplier Diversification}
    \label{benifit3}
\end{figure}

For example, from a few pages, we mapped a subset of the supply chain of nickel, as shown in Figure \ref{benifit3}, identifying who produces nickel, where these companies are located, and what other companies they own. This mapping allows decision-makers to understand the supply chains of specific materials and make informed choices about suppliers.
Understanding such supply chains can significantly aid in making key decisions in choosing the right suppliers. For instance, a decision-maker interested in a specific material can use this framework to identify companies that dominate the supply of that material and gain insights into their operations and dependencies. \appendixautorefname{C} provides a more detailed network illustration and similar examples of mapping different product supply networks.

While this simplified case study serves as a proof of concept, it's important to emphasize that the potential of our framework lies in its scalability and adaptability. The knowledge graph we constructed, though based on a limited set of sources, already reveals complex relationships and valuable insights. However, this is merely scratching the surface of what is possible. 

In real-world applications, our framework can ingest and process vast amounts of data from diverse sources, including industry reports, news feeds, and social media. As more data is incorporated, the knowledge graph grows exponentially richer, uncovering deeper connections and more nuanced insights. Furthermore, the framework's flexibility allows it to be tailored to various industries and supply chain networks beyond electric vehicle manufacturing. Whether applied to pharmaceuticals, semiconductors, or consumer goods, the core principles remain the same – mapping complex relationships, identifying vulnerabilities, and uncovering opportunities.

\section{Conclusions}

This paper has presented a new framework for enhancing supply chain visibility through the integration of knowledge graphs and large language models. Our approach addresses critical challenges in modern supply chain management, particularly in the context of complex, global networks such as those found in the electric vehicle industry.

Our research has shown that this framework can potentially enhance supply chain visibility, offering a more holistic view of intricate supplier networks and material flows. 

Our approach addresses critical challenges in modern supply chain management, particularly in complex, global networks like those in the electric vehicle industry. The key contributions of this work include the development of a scalable methodology for constructing comprehensive supply chain knowledge graphs from diverse, publicly available data sources, and the leveraging of zero-shot large language models to extract and contextualize complex supply chain relationships beyond tier-1 and tier-2 suppliers. We have demonstrated the framework's effectiveness through a case study on electric vehicle manufacturers, revealing critical dependencies and alternative sourcing options for essential minerals and materials. This framework provides decision-makers with insights into supply chain structures, enabling more informed risk management and strategic planning.
While our research has shown significant potential for enhancing supply chain visibility, we acknowledge several limitations. The static nature of the generated knowledge graph fails to capture the dynamic nature of supply chains. Our heavy reliance on publicly available web data may result in incomplete visibility if certain information is not disclosed. Additionally, the reliability of the knowledge graph depends on the credibility of the information sources used, which could lead to inaccuracies if the data is unreliable or outdated. The ability of decision-makers to extract understanding from the framework needs to be validated through further use cases. 
Future work will focus on addressing these limitations. We plan to explore temporal knowledge graph techniques to better represent evolving supply chain relationships over time. We will also enhance our evaluation methodology by developing methods to estimate recall metrics and establish a gold standard for supply chain information extraction. These efforts aim to refine and expand the capabilities of our framework, further improving its utility in supply chain management.

In conclusion, this research represents a significant step in improving supply chain visibility, offering a tool for mapping diverse networks within the supply chain. This work sets the stage for further research to address various use cases, including ethical considerations and sustainability. Moreover, we establish a baseline for the academic community, encouraging further refinement of the framework through enhanced prompting techniques or the integration of few-shot learning tailored to specific supply chain domains.

\section*{Disclosure statement}
No conflict of interest was reported by the author(s).

\section*{Availability of Data}

The data supporting the findings of this study were derived from publicly available Wikipedia pages. These pages were used as resources to feed into the framework for extracting nodes and links. All Wikipedia pages utilized in this study are accessible online.

\bibliographystyle{chicago}
\bibliography{arXiv_Submission}

\newpage

\begin{appendices}
\appendixname{A} \label{AppendixA}
\counterwithin{figure}{section}

\renewcommand{\thefigure}{ A\arabic{figure}}
\setcounter{figure}{0}

\setcounter{section}{0}

\begin{sidewaysfigure} 
    \centering
    \includegraphics[width=\linewidth]{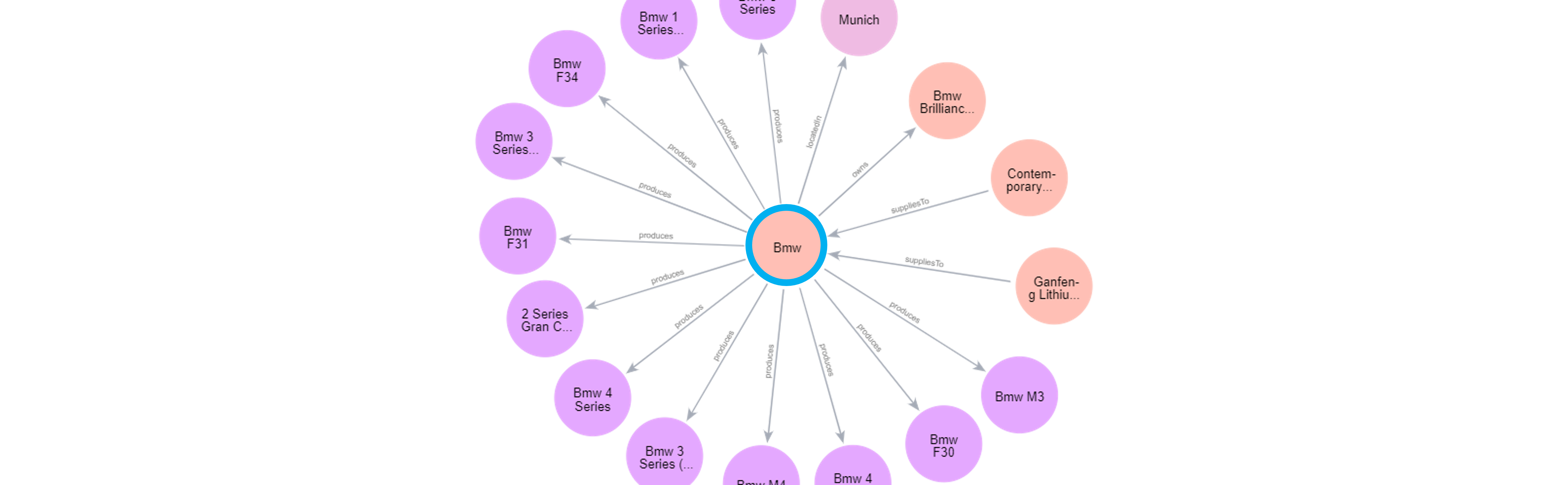}
    \caption{Actual visualization of nodes and links on Neo4j (Extended Network Visibility part-1)}

\end{sidewaysfigure}
\pagebreak

\begin{sidewaysfigure} [p]
    \centering
    \includegraphics[width=\linewidth]{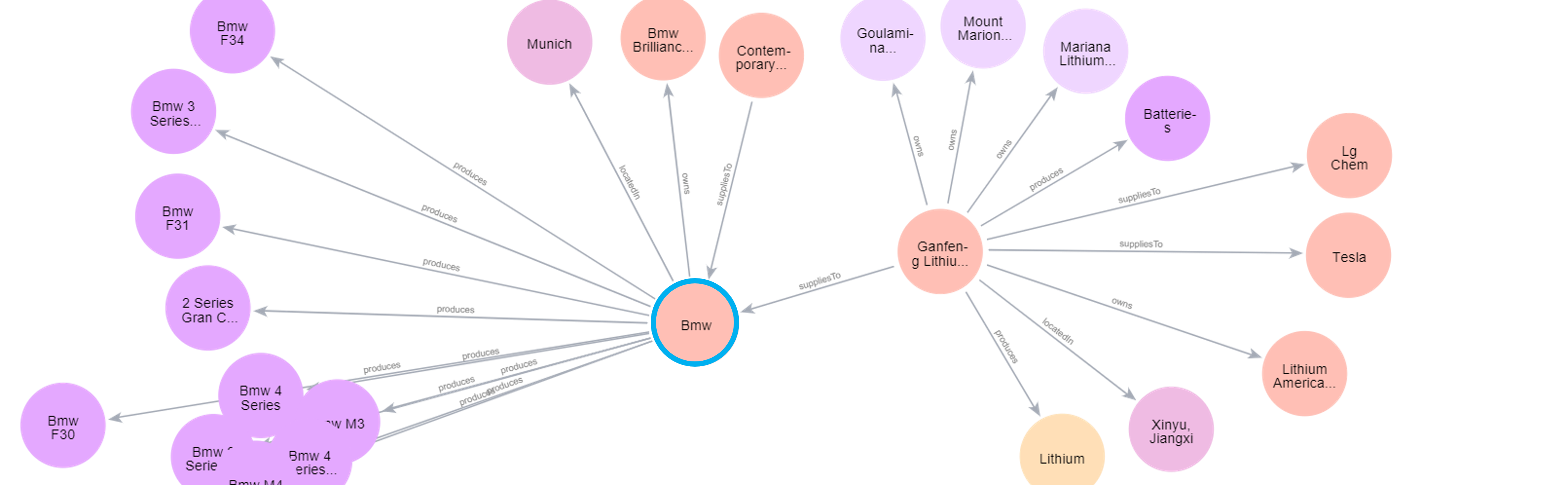}
    \caption{Another visualization of nodes and links on Neo4j (Extended Network Visibility part-2)}

\end{sidewaysfigure}
\pagebreak

\begin{sidewaysfigure} [p]
    \centering
    \includegraphics[width=\linewidth]{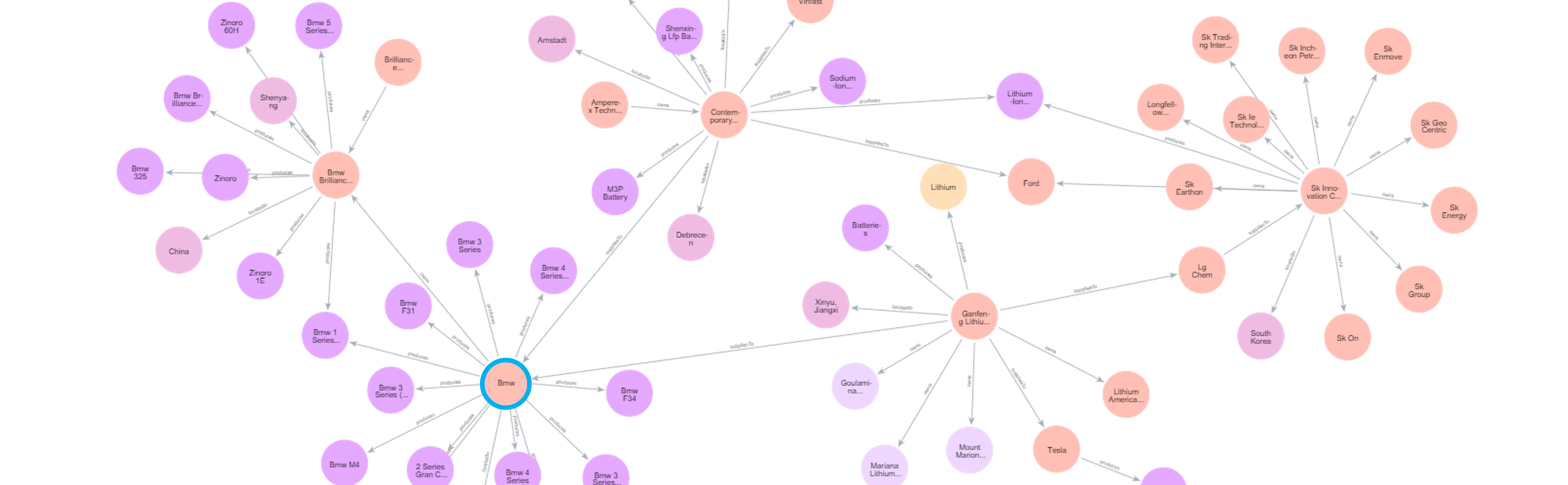}
    \caption{Further visualization of nodes and links on Neo4j (Extended Network Visibility part-3)}

\end{sidewaysfigure}

\pagebreak

\clearpage

\appendixname{B} \label{AppendixA}
\renewcommand{\thefigure}{B\arabic{figure}}
\setcounter{figure}{0}
\begin{sidewaysfigure}[p]
    \centering
    \includegraphics[width=1\linewidth]{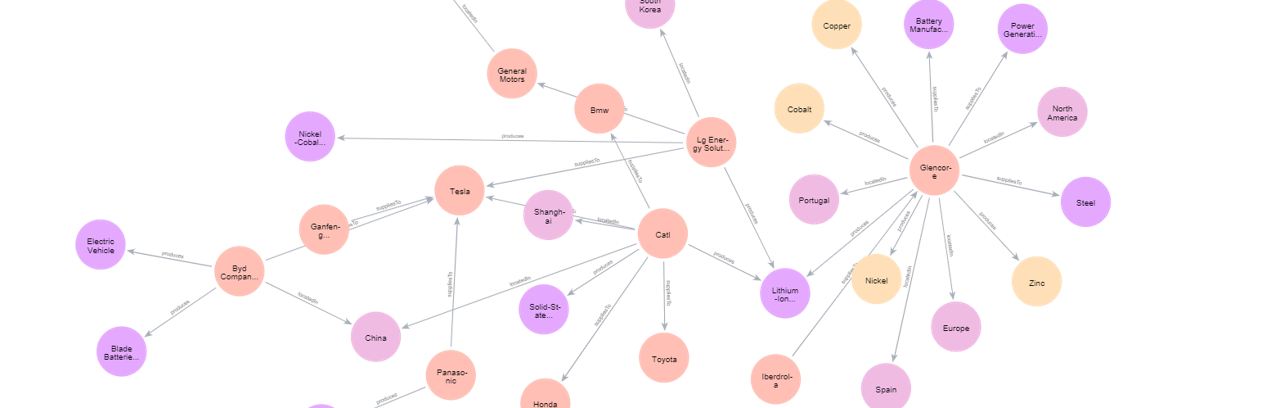}
    \caption{Actual visualization of nodes and links on Neo4j (Alternative Suppliers within the Network)}

\end{sidewaysfigure}
\pagebreak

\clearpage

\appendixname{C} \label{AppendixC} 
\renewcommand{\thefigure}{C\arabic{figure}}
\setcounter{figure}{0}
\setcounter{subsection}{0}
\begin{sidewaysfigure}[p]
    \centering
    \includegraphics[width=1\linewidth]{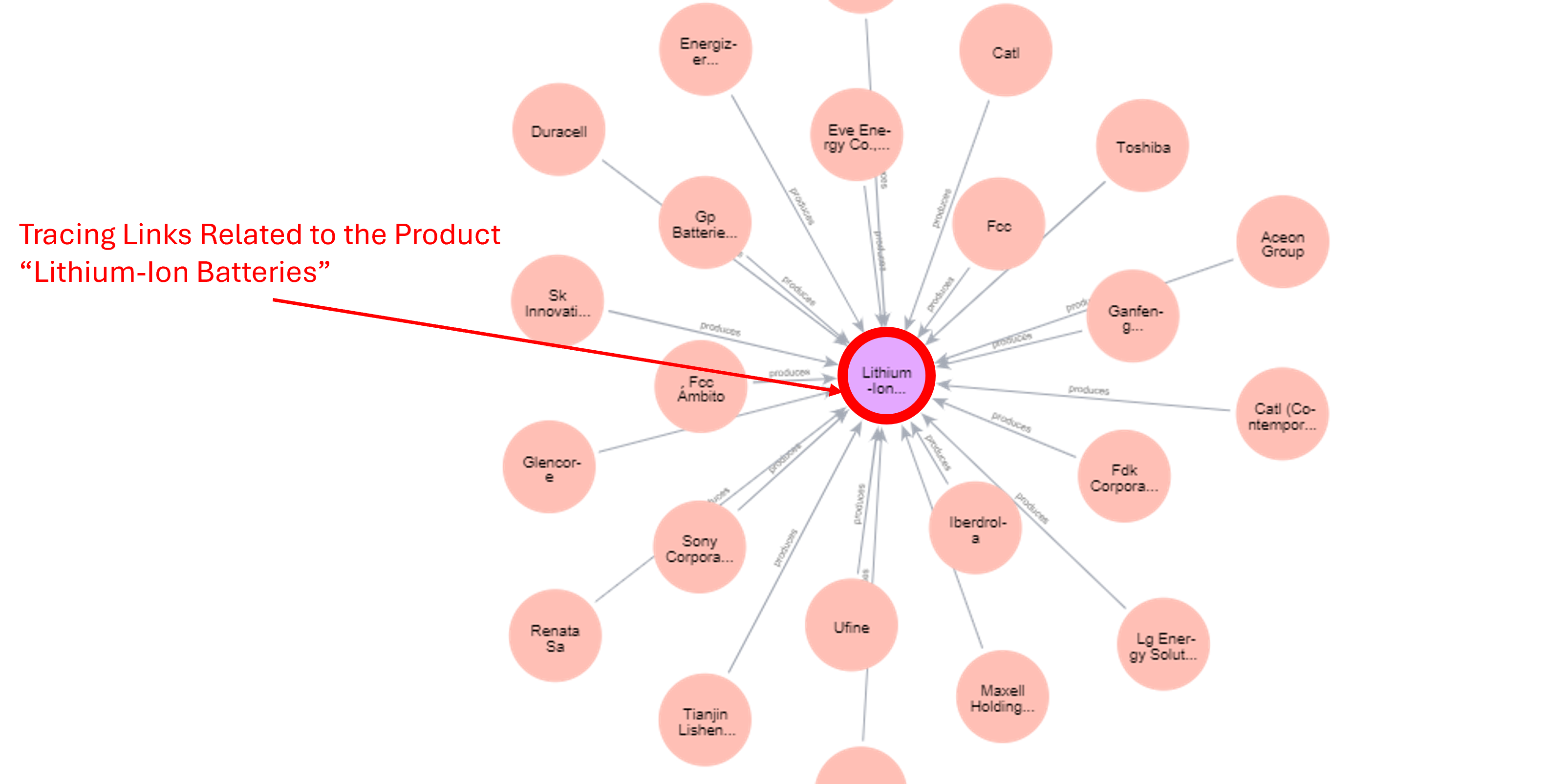}
    \caption{Actual visualization of nodes and links on Neo4j (Tracing Lithium Battery Suppliers and other dependencies (part-1)}

\end{sidewaysfigure}
\pagebreak

\begin{sidewaysfigure}[p]
    \centering
    \includegraphics[width=1\linewidth]{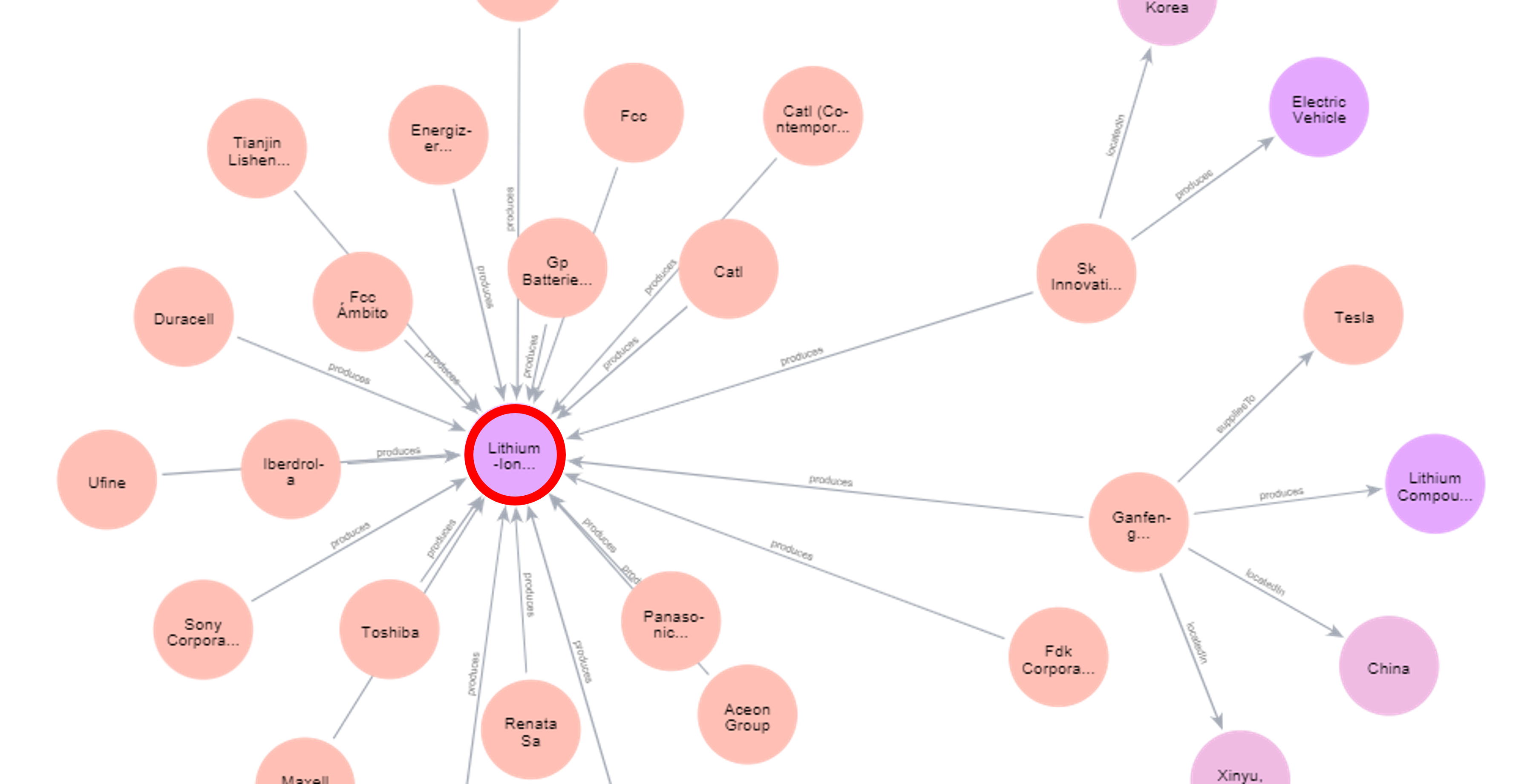}
    \caption{Actual visualization of nodes and links on Neo4j (Tracing Lithium Battery Suppliers and other dependencies (part-2))}

\end{sidewaysfigure}
\pagebreak

\begin{sidewaysfigure}[p]
    \centering
    \includegraphics[width=1\linewidth]{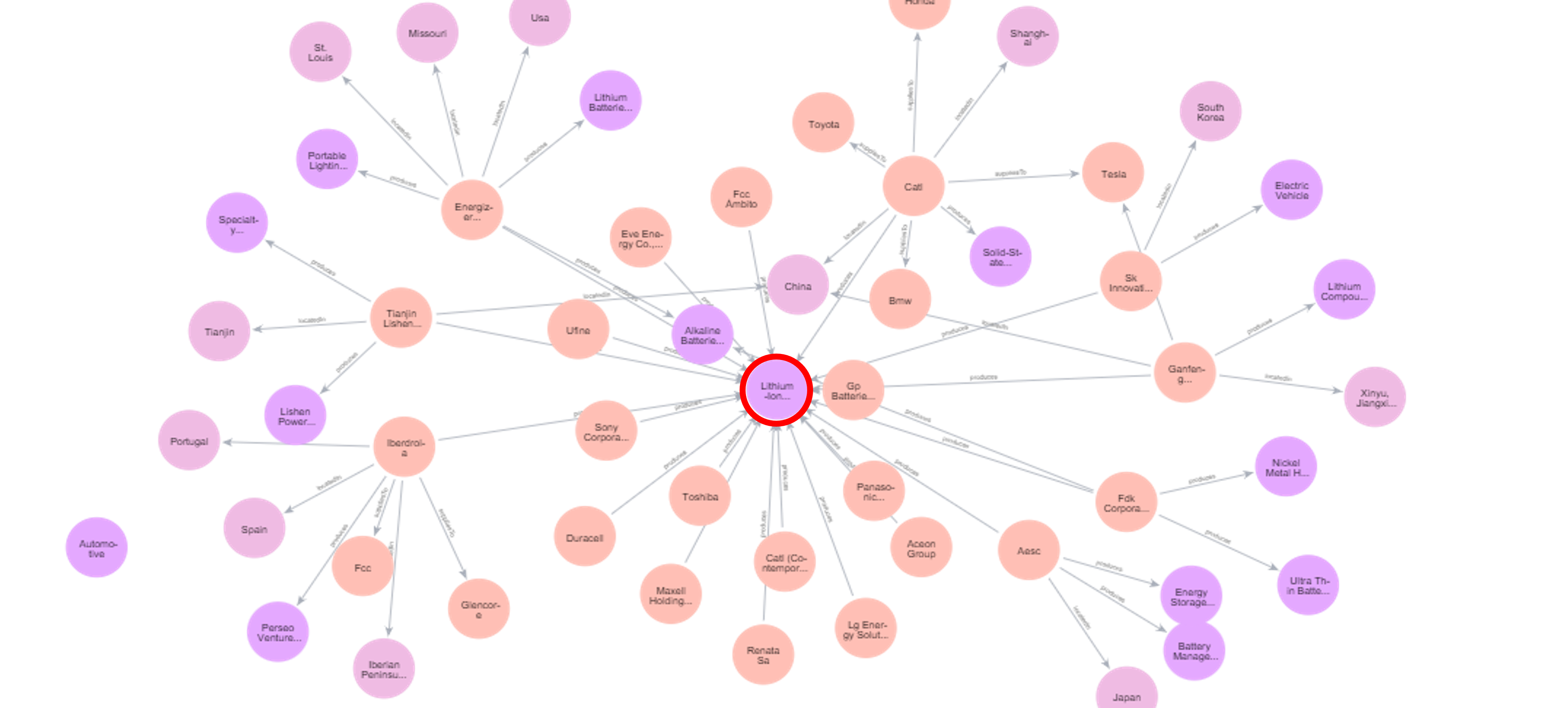}
    \caption{Actual visualization of nodes and links on Neo4j (Tracing Lithium Battery Suppliers and other dependencies (part-3))}

\end{sidewaysfigure}

\end{appendices}
\end{document}